%% file: main.tex
\documentclass[preprint,3p]{elsarticle}
\usepackage[T1]{fontenc}

\usepackage{enumitem}

\usepackage{amsmath,amssymb,amsfonts}
\usepackage[cmintegrals]{newtxmath}
\usepackage{interval}
\usepackage[caption=false,font=footnotesize]{subfig}
\usepackage{balance}
\usepackage{algorithmic}
\usepackage{graphicx,color}
\graphicspath{{./}}
\DeclareGraphicsExtensions{.pdf}
\usepackage{textcomp}
\usepackage[table,xcdraw]{xcolor}
\usepackage{diagbox}
\usepackage{colortbl}
\usepackage{multirow}
\usepackage{multicol}
\usepackage{url}
\usepackage[hidelinks]{hyperref} 
\usepackage[binary-units,per-mode=symbol]{siunitx}
\usepackage{subfig}
\usepackage[nohyperlinks,nolist]{acronym} 
\usepackage[]{algorithm2e}
\usepackage{tikz}

\usepackage{booktabs}
\usepackage{collcell}
\usepackage{hhline}
\usepackage{tabularx}
\newcommand*{\MinNumber}{0.0}%
\newcommand*{\MidNumber}{0.5} %
\newcommand*{\MaxNumber}{1.0}%
\newcommand{\ApplyGradient}[1]{%
        \ifdim #1 pt > \MidNumber pt
            \pgfmathsetmacro{\PercentColor}{max(min(100.0*(#1 - \MidNumber)/(\MaxNumber-\MidNumber),100.0),0.00)} %
            \hspace{-0.33em}\colorbox{green!\PercentColor!yellow}{#1}
        \else
            \pgfmathsetmacro{\PercentColor}{max(min(100.0*(\MidNumber - #1)/(\MidNumber-\MinNumber),100.0),0.00)} %
            \hspace{-0.33em}\colorbox{red!\PercentColor!yellow}{#1}
        \fi
}
\newcolumntype{Y}{>{\centering\arraybackslash}X}
\newcolumntype{L}{>{\arraybackslash}X}
\newcolumntype{R}{>{\raggedleft\arraybackslash}X}
\newcolumntype{C}[1]{>{\centering\arraybackslash}p{#1}}
\newcolumntype{G}[1]{>{\collectcell\ApplyGradient}#1<{\endcollectcell}}
\usepackage[absolute,showboxes]{textpos}
\usepackage{adjustbox}
\usetikzlibrary{patterns}
\usetikzlibrary{decorations}
\usetikzlibrary{spy}
\usepackage{pgfplots}
\usepackage{pgfplotstable}
\pgfplotsset{compat=newest}
\usepgfplotslibrary{units}
\usepgfplotslibrary{groupplots}
\makeatletter
\pgfplotsset{
  /pgfplots/ybar legend/.style={
    legend image code/.code={
      \draw [#1] (0cm,-0.1cm) rectangle (0.15cm,0.2cm);
    };
  }
}
\pgfplotsset{
  /pgfplots/xbar legend/.style={
    legend image code/.code={
      \draw [#1] (-0.1cm,-0.05cm) rectangle (0.2cm,0.1cm);
    };
  }
}
\pgfplotsset{
    minmax/.style={
        mark=|,
        error bars/.cd,
            x dir=plus,
            x explicit,
            error mark=-,
        /pgfplots/table/.cd,
            y=Time,
            x=Min,
            x error expr=\thisrow{Max}-\thisrow{Min},
    },
    avg/.style={
        mark size=2pt,
        /pgfplots/table/.cd,
            y=Time,
            x=Avg,
    }
}
\pgfplotsset{
    legend image with text/.style={
        legend image code/.code={%
            \node[anchor=center] at (0.3cm,0cm) {#1};
        }
    },
}
\pgfplotsset{
    compat=1.11,
    legend image code/.code={
    \draw[mark repeat=2,mark phase=2]
    plot coordinates {
    (0cm,0cm)
    (0.2cm,0cm)        
    (0.4cm,0cm)         
    };%
    }
  }
\pgfplotsset{
 unit code/.code 2 args=
   \begingroup
   \protected@edef\x{\endgroup\si{#2}}\x
}
\tikzset{
  every pin/.style={fill=LightCoreGray,rectangle,rounded corners=3pt,font=\footnotesize},
  small dot/.style={fill=black,circle,scale=0.3}
  }
\pgfdeclaredecoration{stars}{initial}{\state{initial}[width=6pt,next state=star1]
{
}
\state{star1}[width=4pt,next state=gap]
{
    \pgfuseplotmark{star}
  }
\state{gap}[width=4pt,next state=star1]
{
  }
\state{final}
{
    \pgfpathmoveto{\pgfpointdecoratedpathlast}
}  

}

\pgfdeclaredecoration{starstars}{initial}{\state{initial}[width=6pt,next state=star1]
{
}
\state{star1}[width=4pt,next state=star2]
{
    \pgfuseplotmark{star}
}
\state{star2}[width=4pt,next state=gap]
{
    \pgfuseplotmark{star}
  }
\state{gap}[width=4pt,next state=star1]
{
  }
\state{final}
{
    \pgfpathmoveto{\pgfpointdecoratedpathlast}
}  
}
\tikzset{
  dashstar/.style={dash pattern=on 4pt off 4pt,postaction={decorate,decoration=stars}},
  dashstarstar/.style={dash pattern=on 4pt off 8pt,postaction={decorate,decoration=starstars}},
}

\makeatother
\usetikzlibrary{matrix}

\definecolor{CoreGray}{HTML}{BFBFBF}
\definecolor{CoreBlack}{HTML}{333333}
\definecolor{CoreDarkGray}{HTML}{5F5F5F}
\definecolor{CoreBlue}{HTML}{002E7D}
\definecolor{CoreGreen}{HTML}{008000}
\definecolor{CoreGreen2}{HTML}{6AAC8E}
\definecolor{CoreRed}{HTML}{C80000}
\definecolor{CoreYellow}{HTML}{E6AC00}
\definecolor{CoreWhite}{HTML}{FFFFFF}
\definecolor{CoreMagenta}{HTML}{7030A0}
\colorlet{LightCoreGray}{CoreGray!30}
\colorlet{LightCoreBlack}{CoreBlack!20}
\colorlet{LightCoreBlue}{CoreBlue!20}
\colorlet{LightCoreGreen}{CoreGreen!30}
\colorlet{LightCoreRed}{CoreRed!20}
\colorlet{LightCoreYellow}{CoreYellow!20}
\colorlet{LightCoreWhite}{CoreWhite!20}



\input{commands}
\journal{Vehicular Communications}
\begin{document}
\begin{frontmatter}
\setlength{\TPHorizModule}{\paperwidth}
\setlength{\TPVertModule}{\paperheight}
\TPMargin{5pt}
\begin{textblock}{0.8}(0.1,0.02)
     \noindent
     \footnotesize
     If you cite this paper, please use the original reference:
     T. Salomon, L. Maile, P. Meyer, F. Korf, and T. C. Schmidt, ``Negotiating strict latency limits for dynamic real-time services in vehicular time-sensitive networks,'' Vehicular Communications, vol. 57, p.~100~985, Feb. 2026. doi: \href{https://doi.org/10.1016/j.vehcom.2025.100985}{10.1016/j.vehcom.2025.100985}.
\end{textblock}

\title{Negotiating strict latency limits for dynamic real-time services \\ in vehicular time-sensitive networks}

\author[tud,haw]{Timo Salomon\corref{cor1}}
\ead{timo.salomon@haw-hamburg.de}
\author[tue]{Lisa Maile}
\ead{l.t.maile@tue.nl}
\author[haw]{Philipp Meyer}
\ead{philipp.meyer@haw-hamburg.de}
\author[haw]{Franz Korf}
\ead{franz.korf@haw-hamburg.de}
\author[haw]{Thomas C. Schmidt}
\ead{t.schmidt@haw-hamburg.de}
\cortext[cor1]{Corresponding author}
\affiliation[tud]{
  organization={TU Dresden},
  address={Helmholtzstr. 10},
  postcode={01187},
  city={Dresden},
  country={Germany}
}
\affiliation[haw]{
  organization={HAW Hamburg},
  address={Berliner Tor 5},
  postcode={20099},
  city={Hamburg},
  country={Germany}
}
\affiliation[tue]{
  organization={TU Eindhoven},
  address={Postbus 513},
  postcode={5600 MB},
  city={Eindhoven},
  country={Netherlands}
}

\begin{abstract}
  Future vehicles are expected to dynamically deploy in-vehicle applications within a Service-Oriented Architecture (SOA) while critical services continue to operate under hard real-time constraints.
	Time-Sensitive Networking (TSN)  on the in-vehicle Ethernet layer is dedicated to  
   ensure deterministic communication between critical services; its  Credit-Based Shaper (CBS) supports dynamic resource reservations. 
  However, the dynamic nature of service deployment challenges network resource configuration, since any new reservation may change the latency of already validated flows. Standard methods of worst-case latency analysis for CBS have been found incorrect, and current TSN stream reservation procedures lack mechanisms to signal application layer Quality-of-Service (QoS) requirements or verify deadlines.

  In this paper, we propose and validate a QoS negotiation scheme that interacts with the TSN network controller to reserve resources while ensuring latency bounds. 
  For the first time, this work comparatively evaluates reservation schemes using worst-case analysis and simulations of a realistic In-Vehicle Network (IVN) and demonstrates their impact on QoS guarantees, resource utilization, and setup times.
  We find that only one reservation scheme utilizing per-queue delay budgets and network calculus provides valid configurations and guarantees acceptable latency bounds throughout the IVN. 
  The proposed service negotiation mechanism efficiently establishes 450 vehicular network reservations in just \SI[mode=match]{11}{\milli\second}.
\end{abstract}


\end{frontmatter}

\input{acronyms}
\input{1_intro}
\input{2_problem}
\input{3_signaling}
\input{5_networkcalculus}

\input{6_evaluation}
\input{7_conclusion}

\section*{Acknowledgment}
This work was partly supported by the German Federal Ministry of Research, Technology and Space (BMFTR) within the project {AI.Auto-Immune} (16KIS2333 and 16KIS2332K).
During the preparation of this work the authors used artificial intelligence (AI) tools such as Le Chat (\url{chat.mistral.ai}) and ChatGPT (\url{chatgpt.com}) in order to improve conciseness and grammar of parts and original texts written by the authors. After using these services, the authors reviewed and edited the content as needed and take full responsibility for the content of the published article.

\bibliographystyle{elsarticle-num}
\bibliography{bbl/local}

\end{document}

%% file: commands.tex
\usepackage{pifont}

\usepackage{xspace}
\newcommand{\etal}{\textit{et~al.}~}
\newcommand{\eg}{{e.g.,}~}
\newcommand{\ie}{{i.e.,}~}
\newcommand{\cf}{{cf.,}~}

\newcommand{\one}{({\em i})\xspace}
\newcommand{\two}{({\em ii})\xspace}
\newcommand{\three}{({\em iii})\xspace}

\newcommand{\problem}[1]{{\em{P#1}}\xspace}

\usepackage{textcmds}

\newcommand\codeword[1]{\mbox{\textbf{\texttt{\textcolor{CoreBlack}{#1}}}}}
\usepackage{mathtools}

\newcommand\FSI{\mathit{FSI}} 
\newcommand\MFS{\mathit{MFS}} 
\newcommand\MIF{\mathit{MIF}} 
\newcommand\idSL{\mathit{idSl}} 
\newcommand\CMI{\mathit{CMI}} 

%% file: acronyms.tex
\begin{acronym}[DYRECTsn]
	\acro{ACC}[ACC]{Adaptive Cruise Control}
	\acro{ACDC}[ACDC]{Automotive Cyber Defense Center}
	\acro{ACL}[ACL]{Access Control List}
	\acro{AD}[AD]{Anomaly Detection}
	\acro{ADAS}[ADAS]{Advanced Driver Assistance System}
	\acro{ADS}[ADS]{Anomaly Detection System}
	\acroplural{ADS}[ADSs]{Anomaly Detection Systems}
	\acro{ALPN}[ALPN]{Application Layer Protocol Number}
	\acro{API}[API]{Application Programming Interface}
	\acro{AUTOSAR}[AUTOSAR]{\textit{AUTomotive Open System ARchitecture}}
	\acro{AVB}[AVB]{Audio Video Bridging}
	\acro{ARP}[ARP]{Address Resolution Protocol}
	\acro{ATS}[ATS]{Asynchronous Traffic Shaping}
	\acro{BE}[BE]{Best-Effort}
	\acro{C2X}[Car2X]{Car-to-Everything}
	\acro{CA}[CA]{Certification Authority}
	\acroplural{CA}[CAs]{Certification Authorities}
	\acro{CAN}[CAN]{Controller Area Network}
	\acro{CBM}[CBM]{Credit Based Metering}
	\acro{CBS}[CBS]{Credit Based Shaper}
	\acro{CNC}[CNC]{Central Network Controller}
	\acro{CUC}[CUC]{Central User Configuration}
	\acro{CMI}[CMI]{Class Measurement Interval}
	\acro{CoRE}[CoRE]{Communication over Realtime Ethernet}
	\acro{CT}[CT]{Cross Traffic}
	\acro{CM}[CM]{Communication Matrix}
	\acro{DANE}[DANE]{DNS-Based Authentication of Named Entities}
	\acro{DANCE}[DANCE]{DANE Authentication for Network Clients Everywhere}
	\acro{DoS}[DoS]{Denial of Service}
	\acro{DDoS}[DDoS]{Distributed \acl{DoS}}
	\acro{DDS}[DDS]{Data Distribution Service}
	\acro{DNS}[DNS]{Domain Name System}
	\acro{DNSSEC}[DNSSEC]{DNS Security Extensions}
	\acro{DPI}[DPI]{Deep Packet Inspection}
	\acro{DNC}[DNC]{Deterministic Network Calculus}
	\acro{DRL}[DRL]{Deep Reinforcement Learning}
	\acro{DTLS}[DTLS]{Datagram Transport Layer Security}
	\acro{DYRECTsn}[DYRECTsn]{\textit{DYnamic Reliable rEal-time Communication in Tsn}}
	\acro{E/E}[E/E]{Electrical/Electronic}
    \acro{E2E}[E2E]{End-to-End}
	\acro{ECU}[ECU]{Electronic Control Unit}
	\acroplural{ECU}[ECUs]{Electronic Control Units}
	\acro{FN}[FN]{False Negative}
	\acro{FP}[FP]{False Positive}
	\acro{FDTI}[FDTI]{Fault Detection Time Interval}
	\acro{FHTI}[FHTI]{Fault Handling Time Interval}
	\acro{FRTI}[FRTI]{Fault Reaction Time Interval}
	\acro{FTTI}[FTTI]{Fault Tolerant Time Interval}
	\acro{GCL}[GCL]{Gate Control List}
	\acro{HTTP}[HTTP]{Hypertext Transfer Protocol}
	\acro{HMI}[HMI]{Human-Machine Interface}
	\acro{HPC}[HPC]{High-Performance Controller}
	\acro{HSM}[HSM]{Hardware Security Module}
	\acro{IA}[IA]{Industrial Automation}
	\acro{IAM}[IAM]{Identity- and Access Management}
	\acro{ICT}[ICT]{Information and Communication Technology}
	\acro{IDS}[IDS]{Intrusion Detection System}
	\acroplural{IDS}[IDSs]{Intrusion Detection Systems}
	\acro{IEEE}[IEEE]{Institute of Electrical and Electronics Engineers}
	\acro{IGMP}[IGMP]{Internet Group Management Protocol}
	\acro{IoT}[IoT]{Internet of Things}
	\acro{IP}[IP]{Internet Protocol}
	\acro{IVN}[IVN]{In-Vehicle Network}
	\acroplural{IVN}[IVNs]{In-Vehicle Networks}
	\acro{KSK}[KSK]{Key Signing Key}
	\acro{LIN}[LIN]{Local Interconnect Network}
	\acro{MAC}[MAC]{Message Authentication Code}
	\acro{ML}[ML]{Machine Learning}
	\acro{MTU}[MTU]{Maximum Transmission Unit}
	\acro{MOST}[MOST]{Media Oriented System Transport}
	\acro{NAD}[NAD]{Network Anomaly Detection}
	\acro{NADS}[NADS]{Network Anomaly Detection System}
	\acroplural{NADS}[NADSs]{Network Anomaly Detection Systems}
	\acro{NC}[NC]{Network Calculus}
	\acro{OEM}[OEM]{Original Equipment Manufacturer}
	\acro{OMG}[OMG]{Object Management Group}
	\acro{OTA}[OTA]{Over-the-Air}
	\acro{P4}[P4]{Programming Protocol-independent Packet Processors}
	\acro{PCAP}[PCAP]{Packet Capture}
	\acro{PCAPNG}[PCAPNG]{PCAP Next Generation Dump File Format}
	\acro{PCP}[PCP]{Priority Code Point}
	\acro{PKI}[PKI]{Public Key Infrastructure}
	\acro{PSFP}[PSFP]{Per-Stream Filtering and Policing}
	\acro{RAP}[RAP]{Resource Allocation Protocol}
	\acro{RC}[RC]{Rate-Constrained}
	\acro{REST}[ReST]{Representational State Transfer}
	\acro{RPC}[RPC]{Remote Procedure Call}
	\acro{RTPS}[RTPS]{Real-time Publish-Subscribe}
	\acro{SaaS}[SaaS]{Software-as-a-Service}
	\acro{SD}[SD]{Service Discovery}
	\acro{SDN}[SDN]{Software-Defined Networking}
	\acro{SDN4CoRE}[SDN4CoRE]{Software-Defined Networking for Communication over Real-Time Ethernet}
	\acro{SecVI}[SecVI]{\textit{Security for Vehicular Information}}
	\acro{SOA}[SOA]{Service-Oriented Architecture}
	\acroplural{SOA}[SOAs]{Service-Oriented Architectures}
	\acro{SOA4CoRE}[SOA4CoRE]{Service-Oriented Architecture for Communication over Real-Time Ethernet}
	\acro{SOME/IP}[SOME/IP]{\textit{Scalable service-Oriented MiddlewarE over IP}}
	\acro{SPOF}[SPOF]{Single Point of Failure}
	\acro{SR}[SR]{Stream Reservation}
	\acro{SRV}[SRV]{Service}
	\acro{SRP}[SRP]{Stream Reservation Protocol}
	\acro{SVM}[SVM]{Support Vector Machine}
	\acro{SVCB}[SVCB]{Service Binding}
	\acro{SW}[SW]{Switch}
	\acroplural{SW}[SWs]{Switches}
	\acro{TAS}[TAS]{Time-Aware Shaper}
	\acro{TCP}[TCP]{Transmission Control Protocol}
	\acro{TDMA}[TDMA]{Time Division Multiple Access}
	\acro{TN}[TN]{True Negative}
	\acro{TP}[TP]{True Positive}
	\acro{TLS}[TLS]{Transport Layer Security}
	\acro{TSN}[TSN]{Time-Sensitive Networking}
	\acroplural{TSN}[TSN]{Time-Sensitive Networks}
	\acro{TSSDN}[TSSDN]{Time-Sensitive Software-Defined Networking}
	\acro{TT}[TT]{Time-Triggered}
	\acro{TTE}[TTE]{Time-Triggered Ethernet}
	\acro{TTL}[TTL]{Time to Live}
	\acro{UDP}[UDP]{User Datagram Protocol}
	\acro{UN}[UN]{United Nations}
	\acro{QoS}[QoS]{Quality-of-Service}
	\acro{V2X}[V2X]{Vehicle-to-X}
	\acro{WS}[WS]{Web Services}
	\acro{ZC}[ZC]{Zone-Controller}
	\acroplural{ZC}[ZCs]{Zone-Controllers}
	\acro{ZSK}[ZSK]{Zone Signing Key}
\end{acronym}

%% file: 1_intro.tex

\section{Introduction}
\label{sec:introduction}
Vehicles comprise a distributed system of software-defined and hardware-enabled functions.
The \ac{IVN} connects sensors and actuators with in-car intelligence that executes on \acp{ECU} or \acp{HPC}.
With the advent of \acp{ADAS} and autonomous driving, previously isolated domains interconnect at increasing bandwidths.
A central Ethernet backbone is envisioned to soon replace the current \ac{CAN}-based topology~\cite{wtm-avnjr-21,psjtx-svtjr-23,mhlks-fsaad-24}.
Gateways translate between different networks (\eg CAN and Ethernet) and protocols for interoperability and backward compatibility~\cite{iks-aiejr-22}.
For future \acp{IVN}, consolidation with sharing of computational and network resources across applications promises to significantly reduce system complexity and integration cost~\cite{bmle-cssjr-23}.

\subsection{The Software-Defined Car}
The automotive industry is migrating to the \ac{SaaS} paradigm, which enables new business models of greater flexibility, extensibility, and customization~\cite{bmle-cssjr-23}.
Software drives innovation in vehicle performance, safety, and comfort and increasingly contributes to the value of a car~\cite{c-hsijr-21}.
Complementary innovations, such as the online capabilities of connected vehicles, give rise to new software life cycles with frequent updates.

A \ac{SOA} enhances this flexibility through well-defined interfaces, promoting reusability of functions~\cite{bmle-cssjr-23}.
At runtime, service providers announce endpoints, which clients discover via a publish-subscribe model.
Examples of such \textit{dynamic services} include 
context-dependent features (\eg adaptive cruise control, parking assistance),
aftermarket software (\eg infotainment apps, \acp{ADAS}),
and hardware add-ons (\eg trailers with sensors, lights, brakes).

For future deployments, dynamic service orchestration is envisioned, tailored to customer configurations and available resources~\cite{lbdll-eaujr-24,hmks-stsnv-23}.
Post-sale updates and functional upgrades further diversify in-car deployments so that each vehicle may require a unique network configuration based on its active services and hardware capabilities.
These services exhibit a wide range of \ac{QoS} requirements~\cite{chrmk-qaasc-19},  some of which provided by end devices or applications, others demanding network support to address reliability and hard real-time constraints.
Manually pre-calculating network configurations for every possible combination of features, hardware (including legacy vehicles), and post-sale upgrades is impractical for manufacturers. 
Dynamic \ac{IVN} reconfiguration locally computed in the car offers a solution by 
\one adapting the network to active services in real time such that latency limits for data transmissions can be guaranteed, and
\two optimizing resource occupation (and provisioning) by disabling non-critical features when idle.

\subsection{Real-Time Communication in Vehicles}
The \ac{TSN} standards (IEEE 802.1Q~\cite{ieee8021q-22}) offer real-time admission control with ingress filtering, traffic prioritization, and shaping algorithms. 
The \ac{CBS} algorithm~\cite{ieee8021q-22} has been recognized~\cite{mvnb-ipcjr-18,hmks-snsti-19,wtm-avnjr-21} as a promising solution for shaping in-vehicle real-time traffic due to its low complexity.
\ac{CBS} does not require precise time synchronization and allows dynamic bandwidth allocation.
However, all traffic of the same priority class shares the reserved bandwidth, which directly affects worst-case latency and makes it challenging to determine the required bandwidth that guarantees deadlines for each subscription (details in Sec.~\ref{sec:problem}).
For instance, when a new service is added, the reservation of additional subscriptions can change queueing delays for existing ones. 
Since CBS distributes the reserved bandwidth between active subscriptions, previously guaranteed latency bounds may no longer hold, requiring careful admission control and bandwidth reallocation.

For resource reservation and deadline verification in a dynamic \ac{SOA}, the network must identify services, subscriptions, and \ac{QoS} requirements. 
\ac{TSN} operates at the data link layer, complicating integration of service requirements from service discovery protocols on the session layer --  we elaborate on these problems in Sec.~\ref{sec:problem}. 
While \ac{TSN} defines a central user configuration for pre-defined application requirements, the protocol for dynamic signaling of \ac{QoS} requirements between services and the controller remains unspecified~\cite[Section~46.2.2]{ieee8021q-22}.
In previous work, we integrated the automotive service discovery with \ac{SDN}~\cite{hmmsk-dssin-23} to adapt the network to active subscriptions but left the question of signaling and enforcing \ac{QoS} requirements open, which we address in Sec.~\ref{sec:signaling}. 

Traditionally, central controller tools derived static global configurations after all subscriptions were registered. 
Current dynamic approaches do not verify the deadline, and established \ac{TSN} standard formulas for determining worst-case latency for \ac{CBS} have been proven to be incorrect~\cite{maile_decentral_2023}.
Newer approaches allow for dynamic changes and reconfigurations at runtime~\cite{maile_journal_2022,hmks-stsnv-23}, but an integration and evaluation for the automotive use is missing.
Prior work~\cite{mvnb-ipcjr-18,hmks-snsti-19,wtm-avnjr-21,maile_journal_2022,hmks-stsnv-23}, evaluated the resource reservation problem for \ac{TSN}.
Implications of the interaction with application layer protocols of the dynamic automotive \ac{SOA} have not been considered.

\subsection{Contributions}
In this work, we integrate dynamic \ac{QoS} negotiation for an automotive \ac{SOA} with central \ac{TSSDN} control, such that communication latencies strictly comply with the requested deadlines. 
The key contributions of this paper read. 
\begin{enumerate}
	\item We design a signaling mechanism on the \ac{TSSDN} control plane to negotiate real-time requirements for a dynamic \ac{SOA} and adapt the network configuration to all executing services.
    \item We comparatively evaluate reservation schemes w.r.t. maximum service latencies. We identify and eliminate those that fail to strictly guarantee upper time bounds.
    \item We formally verify that requested deadlines for service communication in \ac{TSN} are met by using the \ac{NC} framework based on per-queue delay budgets~\cite{maile_journal_2022}. This approach supports admission control at runtime.
    \item We analyze the implications of our approach for a realistic \ac{IVN} scenario (previously published in~\cite{mhlks-fsaad-24}) via simulations based on the widely used automotive \ac{SOME/IP}. 
    \item We reveal in evaluations that our approach ensures a worst-case end-to-end latency of \SI{1}{\milli\second} per flow while completing signaling for 450 subscriptions in just \SI{11.1}{\milli\second} on the price of up to three times higher resource provisioning than by baseline \ac{TSN} methods.
\end{enumerate}

The remainder of this article is structured as follows.
Sec.~\ref{sec:background} summarizes background on \ac{TSN} and \ac{SOA} in vehicles and discusses related work. 
Sec.~\ref{sec:problem} discusses the dynamic stream reservation problem for \ac{TSN}. In 
Sec.~\ref{sec:signaling}, we present our signaling scheme for service requirements. 
Sec.~\ref{sec:worst_case_latency} applies state-of-the-art methods to determine required bandwidths and validate deadlines.
We evaluate our approach in Sec.~\ref{sec:eval} using simulations of \one a synthetic systematic study and \two a macroscopic study of a realistic \ac{IVN}. 
We discuss the implications and limitations of our findings in Sec.~\ref{sec:discussion}.
Finally, Sec.~\ref{sec:conclusion_and_outlook} concludes with an outlook.

%% file: 2_problem.tex

\section{Background and Related Work}
\label{sec:background}
The automotive industry is undergoing a transformation towards software-defined vehicles, leveraging \ac{SaaS} principles to enhance flexibility, upgradability, and customization~\cite{c-hsijr-21,bmle-cssjr-23}. 
This shift supports context-dependent features (\eg adaptive cruise control, parking assistance),
aftermarket software (\eg infotainment apps, \acp{ADAS}),
and hardware add-ons (\eg trailers with sensors, lights, brakes).
Dynamic subscriptions establish a communication path between publishers and subscribers, with most services subscribed to by multiple clients.

This evolution, however, also increases \ac{IVN} complexity due to the growing number of optional features and system variants~\cite{c-hsijr-21,bmle-cssjr-23,hmks-stsnv-23,lbdll-eaujr-24}. 
To address this, future \acp{IVN} will likely transition from legacy \ac{E/E} architectures with function-specific \acp{ECU} toward centralized computing and communication platforms~\cite{bmle-cssjr-23,lbdll-eaujr-24}. 
A high-speed Ethernet backbone, replacing traditional \ac{CAN} bus systems, fosters a flat topology with a standardized IP stack~\cite{bmle-cssjr-23,hmks-stsnv-23}.

Middleware solutions are crucial for enabling service discovery, connection setup, and \ac{QoS} provisioning in these agile networks.
They manage the communication channel, which can be either unicast or multicast, on the application layer in form of a subscription, connection, communication flow, or stream -- which we use interchangeably in this paper.
Notable candidates include \ac{DDS}~\cite{o-dsjr-22}, widely used in robotics, and \ac{SOME/IP}~\cite{a-spsjr-24,a-ssdjr-24}, optimized for automotive applications due to its low complexity and overhead. 
Both support runtime service discovery and event-driven publish-subscribe models over UDP or TCP~\cite{a-spsjr-24,o-dsjr-22}. 
We focus on \ac{SOME/IP} due to its widespread adoption in the automotive industry and its selection by the \ac{AUTOSAR} consortium~\cite{a-spsjr-24, iks-aiejr-22}. 
Nonetheless, the proposed approach can be seamlessly adapted to \ac{DDS}. 

Automotive services have diverse \ac{QoS} requirements that the network must guarantee~\cite{chrmk-qaasc-19}.
Dynamic services with frequent updates may change their communication behavior which complicates this task as it breaks static real-time configurations, which are common in current \ac{IVN} setups~\cite{mvnb-ipcjr-18,mhlks-fsaad-24}.
On the other hand, dynamic configurations come with challenges, which we will explain in Sec.~\ref{sec:problem}.

\subsection{Shaping Time-Sensitive Traffic in Cars}
\label{sec:background_tsn}
The use of TSN in vehicles has recently gained significant attention~\cite{psjtx-svtjr-23}.
The automotive profile (draft IEEE 802.1DG-2024~\cite{ieee8021dg-24}) outlines its application in cars.
The TSN standards under the IEEE 802.1Q umbrella~\cite{ieee8021q-22} define building blocks for real-time admission control.
Prioritization reduces high priority traffic latency, whereas shaping algorithms reduce jitter and prevent starvation of lower priorities by limiting the bandwidth for high priority traffic.

This work focuses on egress traffic shaping to ensure \ac{E2E} latency guarantees. 
A VLAN \ac{PCP} maps to (usually 8) strict priority queues.
The \ac{TAS} (IEEE 802.1Qbv) adds a transmission selection algorithm and a gate to each queue.
\ac{TDMA}-scheduled gates can minimize latency and jitter but require precise synchronization and a complex, typically offline-computed schedule~\cite{som-ersjr-24}. 
In contrast, \textit{asynchronous} shaping algorithms such as \ac{CBS} (IEEE 802.1Qav) or \ac{ATS} (IEEE 802.1Qcr) manage bandwidth allocation without precise time synchronization. 

\ac{TDMA} schedules for \ac{TAS} usually involve all endpoints and switches to implement a synchronized, network-wide schedule. 
Therefore, they often assume predefined applications, making them unsuitable for dynamic services. 
While dynamic scheduling techniques exist~\cite{gavrilut_avb-aware_2018}, runtime modifications remain challenging.
Our previous investigations for updating \ac{TAS} \ac{TDMA} schedules~\cite{hmks-stsnv-23} revealed the need for precise timing of transactional updates that ensure reconfigurations do not introduce permanent delays. 
Here, asynchronous mechanisms have a distinct advantage, as they do not rely on precise timings and can be updated at runtime.
A hybrid approach is emerging where some flows remain predefined with static schedules, while flows for dynamic services rely on shaping mechanisms that support dynamic updates, such as \ac{CBS}. 
Leonardi \etal suggest partitioning \ac{TSN} queues to isolate dynamic from static priorities~\cite{lbp-bptjr-21}.

\ac{ATS} offers per-stream traffic shaping based on a token bucket algorithm that performs well for sporadic traffic~\cite{ntaws-pcijr-19}, possibly achieving lower \ac{E2E} latencies than \ac{CBS}~\cite{flgx-sacjr-20, zlbpy-sttjr-21}, making it a promising candidate for future \acp{IVN}.
However, Thomas \etal demonstrate that \ac{ATS} can cause unbounded latencies for specific traffic patterns in various networks~\cite{tl-ncsjr-24}, including star topologies.
We explored placement and configuration methods for ATS schedulers to prevent this behavior~\cite{lmhks-atsra-25}, which still require mathematical proof for their application in centralized configuration of critical traffic. 
Furthermore, evaluation is currently not possible because there are no frameworks available for dynamically determining ATS configurations with their guaranteed maximal latencies.
In the past, we developed the \ac{DYRECTsn}~\cite{dyrectsn} framework which allows dynamic flow reservation with guaranteed latencies supporting \ac{CBS}.
Thus, while our signaling scheme would support \ac{ATS}, we do not consider it in this work and instead focus on \ac{CBS}.

\ac{CBS} enforces idle times between high-priority transmissions to mitigate negative impact on lower priority traffic.
It is often used due to its low implementation complexity and support for dynamic reservations.
The \codeword{idle slope} queue parameter defines the guaranteed bandwidth and imposes an upper transmission limit. 
The total idle slopes of \ac{CBS} queues at a port should stay below the link rate, leaving some bandwidth for best effort traffic.
Unlike per-flow reservation schemes, \ac{CBS} operates at queue granularity, which simplifies configuration but shifts the complexity to idle slope calculation -- a challenge outlined in Sec.~\ref{sec:problem:dynamic_reservation}.

\begin{figure}
    \centering
    \includegraphics[width=.7\linewidth, trim={20pt 24pt 24pt 20pt}, clip=true]{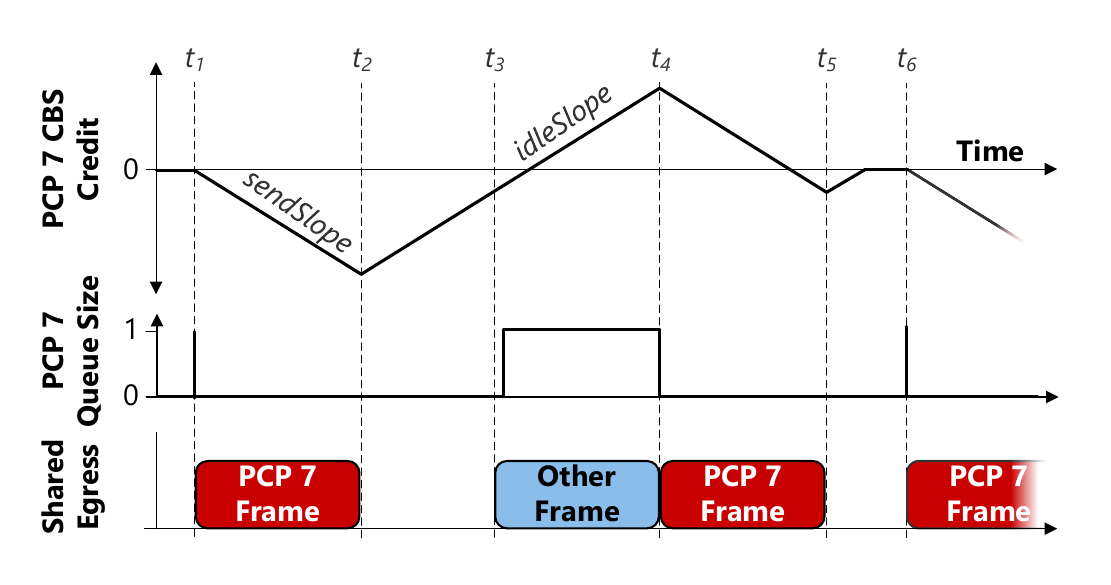}
    \caption{Credit evolution for \ac{CBS} forwarding.}
    \label{fig:cbs}
\end{figure}

\ac{CBS} maintains a \codeword{credit} to determine frame transmission eligibility as shown in Fig.~\ref{fig:cbs}. 
Frames can only begin transmission if the credit is non-negative ($\ge 0$). 
Each transmission decreases the credit at the rate \codeword{send slope} (with $\mathit{send\ slope} = \mathit{idle\ slope} - \mathit{link\ rate}$, \cf $\interval{t_1}{t_2}$, $\interval{t_4}{t_5}$). 
When the credit is negative, it increases at the idle slope rate until reaching zero (\cf $\interval{t_2}{t_3}$, $\interval{t_5}{t_6}$) and lower priorities are allowed to transmit frames (\cf $\interval{t_3}{t_4}$). 
If other priority queues delay an eligible transmission (\cf $\interval{t_3}{t_4}$), the credit can increase above zero at the idle slope rate. 
If a queue is empty and the credit is positive, it resets to zero.

Walrand \etal show key principles for in-vehicle Ethernet architectures using TSN with CBS~\cite{wtm-avnjr-21}: 
\one Minimizing the number of hops reduces chains of influence, thereby decreasing latencies. 
\two \qq{A fast network transport slow flows for free,} so using different link speeds from \SI{10}{\mega\bit\per\second} for control loops to \SI{10}{\giga\bit\per\second} in the backbone improves latency and buffer utilization.

\ac{ML} has transformed various domains and has recently been applied to computer networks~\cite{kkw-mlcjr-22}, including \ac{SDN} and \ac{TSN}.
Laclau \etal apply spectral clustering to predetermine valid network configurations for in-vehicle service setups that can be loaded at runtime~\cite{lbdll-pncjr-23}.
However, their approach does not validate latency bounds and focuses on predetermined configurations rather than fully dynamic adaptation of the network to running services.
Recent studies have demonstrated the application of \ac{DRL} for traffic classification, flow-aware scheduling, and routing~\cite{hzdxy-defjr-23,cl-dnsjr-23,cyzz-rrsjr-25}.
These approaches primarily focus on \ac{TDMA} configuration for the \ac{TAS}. 
In contrast, we focus on dynamic asynchronous mechanisms for shaping dynamic services. 
While these studies show promising directions, they currently lack the formal validation required for ensuring deterministic latency bounds, which is critical for \acp{IVN}. 
Our work aims to apply more rigorous and validated approaches to network adaptation and performance optimization evaluating their practical applicability using the proposed service negotiation method for \ac{TSSDN}. 

Previous CBS deployments in cars with guaranteed latency remained static and neither addressed dynamic service discovery nor verified latency bounds for service configurations, which we address in this work.
We utilize \ac{CBS} for shaping dynamic services as it is widely recognized for in-vehicle deployment~\cite{mvnb-ipcjr-18,hmks-smsdn-19,wtm-avnjr-21}. 
For pre-defined communication we also consider \ac{TAS} with a static \ac{TDMA} schedule.
Our signaling scheme can also be extended to other algorithms.

\subsection{Signaling Quality-of-Service Requirements}
In \ac{TSN}, the \ac{SRP} (IEEE 802.1Qat) announces talker and listener information at runtime, supporting a distributed approach, where switches calculate bandwidth independently, and a recently added centralized approach (IEEE 802.1Qcc), where a controller coordinates the reservation.
The distributed admission control has a high communication overhead compared to a centralized architecture~\cite{ampbp-cacjr-20}.
The upcoming \ac{RAP} (draft IEEE 802.1DD~\cite{ieee8021dd-draft}) will enhance \ac{QoS} provisions, including redundancy. 

The centralized approach is often referred to as \ac{TSSDN}~\cite{ndr-tssdn-16,hmmsk-dssin-23} as it combines TSN shaping with central network control of \ac{SDN}.
Initially used in campus and data center networks~\cite{mabpp-oeicn-08}, \ac{SDN} aims to enhance network control and adaptability~\cite{krvra-sdncs-15}.
In cars, \ac{TSSDN} promises greater flexibility, adaptability, robustness, and security compared to traditional networks~\cite{hmg-rsarn-18,hhlng-saeea-20,hmks-stsnv-23}.
In previous work~\cite{hmks-stsnv-23}, we showed that \ac{TSSDN} allows for adaptable real-time configurations that can at the same time improve network security in cars with strict flow control.

In TSN's central configuration model, a \ac{CUC} can provide predefined application requirements to the controller~\cite{ieee8021q-22}.
Still, protocols are missing that enable applications to signal service requirements, including deadlines, to the control plane.
We address this gap by integrating dynamic \ac{QoS} negotiation with the automotive service discovery.

Automotive \ac{SOA} protocols, such as \ac{SOME/IP}~\cite{a-spsjr-24} or \ac{DDS}~\cite{o-dsjr-22}, operate on the session layer, using the UDP-IP stack without link-layer real-time guarantees. 
Higher layer \ac{QoS} can sometimes be configured, \eg specifying update rates or retransmissions~\cite{o-dsjr-22}.
Mapping \ac{DDS} service requirements statically to \ac{TSN} using the \ac{CUC} has been demonstrated~\cite{lzhzh-tddjr-23}.
However, integrating dynamic TSN stream reservation with higher layer \ac{SOA} is challenging due to the lack of standardized \ac{QoS} translation across OSI layers.

\ac{SDN} enhances network control and adaptability~\cite{krvra-sdncs-15}, optimizing protocols such as the Address Resolution Protocol (ARP) with central network knowledge.
In a \ac{SOA} it can enable service deployment across the infrastructure~\cite{fggc-sdnjr-22}.
Nayak \etal\cite{natgw-rasjr-23} propose a \textit{P4} programmable data plane implementation for SOME/IP, learning subscriptions in network devices.
In previous work, we established the controller as a rendezvous point for \ac{SOME/IP} service discovery~\cite{hmmsk-dssin-23} and related work propose integrations for \ac{DDS}~\cite{bhmba-dbcjr-14}.
Central SDN control can make networks more robust against intruders -- especially in real-time systems, where resource theft becomes a safety issue -- allowing only specific services to communicate~\cite{hmks-stsnv-23}. 
However, signaling \ac{QoS} requirements for dynamic services remains unaddressed.

Previously, we proposed a dynamic \ac{QoS} negotiation protocol utilizing a heterogeneous protocol stack that falls back to \ac{SRP} for resource reservation~\cite{chrmk-qaasc-19}.
Such a multi-stage procedure is common and involves first setting up the subscription in the \ac{SOA}, then establishing a layer 2 subscription for resource reservation using, \eg the \ac{SRP}. 
Coordinating the two subscriptions is error-prone, can cause inconsistencies, and requires workarounds on end devices to bridge OSI layers.
Our service discovery scheme for \ac{TSSDN} introduces a single-stage procedure, including \ac{QoS} signaling and resource reservation with guaranteed latency.

\subsection{Delay Bounds in Time-Sensitive Networking}
Traditional configuration tools generate static global configurations after all subscriptions are registered~\cite{laursen_routing_2016,pop_design_2016,b-adnjr-17,gavrilut_avb-aware_2018,mvnb-ipcjr-18, berisa_avb-aware_2022}. 
New tools offer centralized real-time flow planning and network configuration for dynamic traffic~\cite{maile_journal_2022,ieee8021q-22, hmks-stsnv-23}. 
Therefore, they leverage central knowledge of network topology, active flows, and \ac{QoS} requirements.
Our integrated dynamic \ac{QoS} negotiation utilizes centralized dynamic reservation approaches to guarantee latency bounds.

Extensive work has been devoted to the analytical latency analysis of \ac{TSN}. 
\ac{NC} has been used to analyze \ac{CBS} delays~\cite{queck, zhao_latency_2020}, with comprehensive \ac{NC} results for \ac{TSN} provided in~\cite{maile_network_2020, deng_survey_2022}.
Other methods, such as busy-period analysis, have also been applied to \ac{CBS}~\cite{diemer_modeling_2012, bordoloi_schedulability_2014} and to systems combining \ac{TAS} and \ac{CBS}~\cite{ashjaei_schedulability_2017}. 

The \ac{SRP} has limitations in scheduling a large number of slow flows when relying on a fixed \ac{CMI} for bandwidth allocation~\cite{mvnb-ipcjr-18}.
Furthermore, idle slope calculations as defined in TSN standards do not consistently yield the expected delays~\cite{ashjaei_schedulability_2017}.
These calculations can be overly pessimistic in some cases while failing to provide safe delay guarantees in others~\cite{maile_decentral_2023,zhao_improving_2021}.
The idle slopes according to TSN standards do not always yield expected delays~\cite{ashjaei_schedulability_2017} and can be both overly pessimistic and yet provide un-safe delay guarantees~\cite{maile_decentral_2023,zhao_improving_2021}.
Boiger~\cite{b-adnjr-17} presented a counterexample to the standard's statement~\cite{BA} on maximum delay in \ac{CBS} networks, demonstrating that the \SI{2}{\milli\second} \ac{E2E} delay over seven \SI[per-mode=symbol]{100}{\mega\bit\per\second} hops can be violated. 

In previous work, we proposed a central admission control scheme for dynamic delay-guaranteed flow reservations in CBS networks~\cite{maile_journal_2022}. 
Our \ac{DYRECTsn}~\cite{dyrectsn} framework determines paths, assigns flow priorities based on deadlines, and determines required idle slopes per queue based on delay budgets. 
These delay budgets ensure that the worst-case delay of a flow passing through this queue stays below a maximum latency (budget) even when other flows are added.

The delay budget approach applied in this work (details in Sec.~\ref{sec:worst_case_latency}) requires suitable per-hop latency limits (budgets). 
If these limits are too small, fewer flows can be scheduled; if too high, they allow a higher maximum \ac{E2E} latency and thus increase the smallest deadline that can be guaranteed.
We previously introduced a meta-heuristic approach for optimizing these budgets
For the optimization of the delay budgets, we previously introduced a meta-heuristic method~\cite{dyrectsn}. 
Recently, an approach based on \ac{ML} has been explored~\cite{gswhh-maljr-21}, but it remains limited: it assumes identical budgets at each hop and supports only linear and tree topologies. 
Due to these constraints, we rely on the more flexible meta-heuristic for our \ac{IVN} scenario (Sec.~\ref{sec:eval_car}), which yields tighter network configurations as it allows for individual delay budgets at each hop.

A comparison is missing between the established standard solution and the delay budget approach regarding resource allocation efficiency and latency guarantees for dynamic services.
Additionally, determining delay bounds, signaling QoS requirements, and configuring the network are usually investigated separately.
In this work, we integrate dynamic \ac{QoS} negotiation for an automotive \ac{SOA} with dynamic real-time traffic management comparing the TSN standard stream reservation scheme with the delay budget approach.
We build a comprehensive framework to evaluate the impact of combining dynamic service discovery with different resource reservation schemes gaining valuable insights in realistic automotive scenarios, which differentiates our work from previous studies~\cite{ndr-tssdn-16,mvnb-ipcjr-18,gkbh-sdfjr-19,hmks-snsti-19,hhlng-saeea-20,wtm-avnjr-21,maile_journal_2022,hmks-stsnv-23}.

\section{The Problem of Deadline-Compliant Operation of Dynamic Services}
\label{sec:problem}
We identify two areas with three key challenges in providing real-time guarantees for the dynamic in-vehicle SOA using \ac{CBS}: \one in integration of service discovery and resource allocation, and \two in determining the required idle slopes for dynamically changing service configurations.

\subsection{QoS Signaling for Dynamic Services in TSN}
\label{sec:problem:soa_tsn}
Typical publish-subscribe implementations consist of a \textit{discovery phase} (request, advertise), a \textit{subscription phase} (subscribe, acknowledge), and subsequent data transfer~\cite{a-spsjr-24,o-dsjr-22}.
Some protocols incorporate \ac{QoS} settings in advertisements and subscriptions, others lack QoS negotiation and require additional protocols as proposed in~\cite{chrmk-qaasc-19}.
However, these negotiations remain transparent to the network, necessitating a separate signaling mechanism for resource allocation.

\begin{figure}%
    \centering%
    \subfloat[]{\includegraphics[width=0.6\linewidth,trim={24pt 22pt 18pt 22pt}, clip=true]{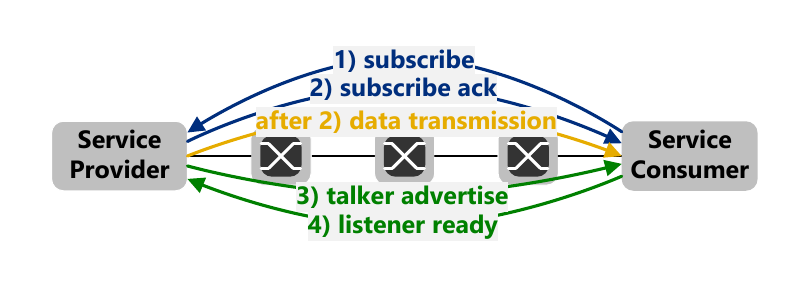}\label{fig:architecture_tsn_decentral}}%

    \subfloat[]{\hspace{0.19\linewidth}\includegraphics[width=0.6\linewidth,trim={24pt 22pt 18pt 24pt}, clip=true]{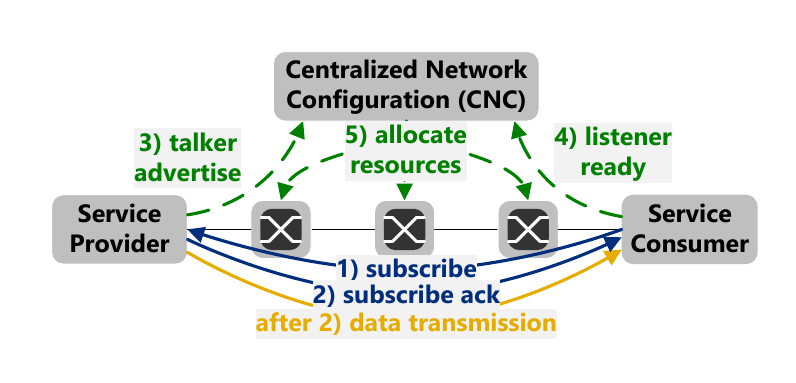}\label{fig:architecture_tsn_central}\includegraphics[width=0.19\linewidth]{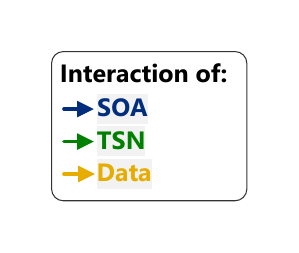}}
    
    \caption{Interaction of separate \textcolor{CoreBlue}{SOA} and \textcolor{CoreGreen}{TSN} subscriptions following a two-stage procedure.
    Subsequent \textcolor{CoreYellow}{data transmission} can start after the SOA subscription.
    (a) Uses distributed stream reservation in each switch; (b) shows the centralized model of TSN, where signaling of application requirements to the control plane remains unspecified (\textcolor{CoreGreen}{dashed}).
    }
    \label{fig:architecture_tsn}
\end{figure}

Fig.~\ref{fig:architecture_tsn_decentral} illustrates such a two-stage procedure.
First, the SOA middleware discovers the service endpoint and establishes a subscription.
Then, for instance, the TSN SRP allocates resources.
The SRP also defines a centralized model (see Fig.~\ref{fig:architecture_tsn_central}), which enables advanced resource allocation algorithms based on central network knowledge.
However, the protocol for signaling application requirements between the data and control plane is unspecified in TSN~\cite[Section~46.2.2]{ieee8021q-22}.
Both approaches share the following problems.

\begin{description}
    \item [\textbf{\problem{1}}:] \textit{Separate SOA and TSN subscriptions can be error prone} 
    as both subscriptions must be coordinated, including timeouts and retries, which may lead to inconsistencies. 
    There is no standardized process for communicating QoS requirements from the application to the network that bridges the OSI layer gap from session layer 5, where the SOA middleware operates, to the data link layer 2, where TSN forwards and shapes the traffic.
\end{description}

\problem{1} creates critical challenges in real-world implementations, primarily revolving around cross-layer coordination, error handling, and resource management.
In practice, applications must either extend QoS configurations via abstractions (e.g., socket API) or bypass them entirely to initiate layer-2 reservations -- often requiring low-level details (e.g., MAC addresses) unavailable to higher layers.
Current OS environments (e.g., Linux) lack support for such vertical integration, forcing applications to manage cross-layer dependencies without robust OS assistance.

Error handling further complicates the implementation of the two-stage procedure. 
SOA subscriptions may succeed while both TSN talker advertisement and listener readiness may fail independently due to errors or resource constraints, potentially disrupting other same-priority flows without additional safeguards.
Ungraceful terminations (\eg crashes) pose further difficulties, where the two subscriptions may expire at different times, leading to duplicate reservations rather than proper state cleanup, thus demanding distinct and carefully coordinated recovery mechanisms. 
Misaligned renewal cycles between the two protocols further complicate synchronization and increase risks of partial connectivity or orphaned reservations.
These issues collectively undermine system reliability and resource efficiency in time-sensitive communication scenarios.

\begin{description}
    \item [\textbf{\problem{2}}:] \textit{Access control is limited to the \ac{SOA} subscription phase}, 
    leaving publishers unable to restrict access to TSN multicast streams. 
    With \ac{SRP}, subscribers can join without publisher involvement, conflicting with automotive service discovery, where publishers must first approve subscribers to enforce access control. 
\end{description}

\problem{2} exposes gaps in \ac{IVN} security mechanisms for access control and authentication.
While the \ac{SOA} can enforce access policies, these are not applied at layer 2.
This is particularly problematic for multicast subscriptions, which account for approximately 75\% of all in-vehicle control flows~\cite{hmks-stsnv-23}. 
TSN multicast is not strictly limited to valid subscribers but to any entity joining the group address.

\subsection{Challenges in Dynamic Reservation}
\label{sec:problem:dynamic_reservation}
\begin{figure}
    \centering
    \hfill
    \subfloat{\includegraphics[width=.3\linewidth, trim={20pt 8pt 20pt 18pt}, clip=true]{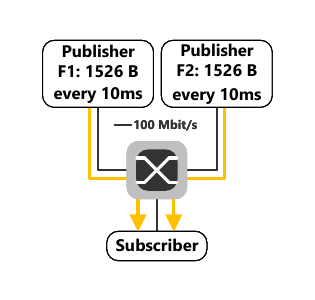}\label{fig:motivation_network}}
    \hfill
    \trimbox{0pt 3pt 2pt 14pt}{\noindent\subfloat{\input{fig3_motivation.tex}\label{fig:delay_bound}}}
    \hfill
    \vspace{-2pt}
    \caption{Network with two flows using \ac{CBS} and the analytical bound of F1 with and without F2 for various idle slopes at the switch queue. 
    Bandwidth depends on the observed interval, so even when senders comply with their announcements, interference from cross traffic can cause sporadic traffic buildup, leading to higher latency.
    }
    \label{fig:motivation}
\end{figure}

Adding flows dynamically to CBS configurations necessitates adjusting idle slopes at run time.
This can impact already validated reservations, complicating dynamic traffic management. 
Fig.~\ref{fig:motivation} illustrates a simple scenario with two flows from different senders directed through a switch to one receiver. 
Both flows share the highest priority, and a \ac{CBS} is configured at the switch queue towards the receiver with an idle slope that limits the bandwidth.
Each sender transmits one maximum Ethernet frame every \SI{10}{\milli\second}, resulting in approx. \SI{1.2}{\mega\bit\per\second} per flow bandwidth.

Fig.~\ref{fig:motivation} also shows the worst-case \ac{E2E} latency for flow F1, determined by static analysis with \ac{DNC}~\cite{LUDBFF}.
The minimum required idle slope of the \ac{CBS} queue equals the actual flow bandwidth. 
When only F1 is active, the delay bound is \SI{0.4}{\milli\second}.
When both F1 and F2 are active the delay bound of F1 rises to \SI{5.2}{\milli\second} at the minimum required idle slope.
Despite senders adhering to their reservations, it is possible that traffic accumulates, and queues grow, resulting in increased latency~\cite{b-adnjr-17,maile_decentral_2023} (\cf Sec.~\ref{sec:eval}).
Increasing the idle slope reduces the queueing delay for F1, \ie achieving a deadline below \SI{1}{\milli\second} requires roughly seven times the active flow bandwidth.

Migge~\etal\cite{mvnb-ipcjr-18} demonstrate drawbacks of the standard reservation scheme, mainly related to the fixed sending interval, and illustrate how it hinders schedulability in realistic automotive networks. 
For flows with large send intervals and small deadlines, significantly more bandwidth must be reserved than the actual flow bandwidth. 
Besides, increased reservation can reduce available resources for lower priorities if fully utilized~\cite{maile_low_priority_2024}. 
This leads us to a third problem:
\begin{description}
    \item [\textbf{\problem{3}}:] 
    \textit{Determining the required idle slopes for dynamic services is non-trivial}. Over-reservation risks resource starvation of lower priorities, and standard procedures do not yield deadline-compliant reservations~\cite{maile_decentral_2023}. 
\end{description}

\problem{3} causes different reservation schemes to degrade latency for both \ac{CBS}-shaped and lower-priority traffic.
The impact of novel approaches for dynamic reservations on a full \ac{IVN} is unknown, which warrants detailed evaluation of reservation schemes.
This is underlined by the current draft for TSN \textit{Shaper Parameter Settings for Bursty Traffic Requiring Bounded Latency}~\cite[Section 3.4]{ieee8021qdq-24}, which states that it is non-trivial to determine accumulated latency for CBS and outside of the scope of this standard.

%% file: fig3_motivation.tex
\begin{tikzpicture}
    \begin{axis}[
        height=.35\linewidth,
        width=.60\linewidth,
        ylabel={Delay bound},
        xlabel={Idle slope},
        y unit=ms,
        ymin=0,
        ymax=0.0054,
        ytick={0.001, 0.002, 0.003, 0.004, 0.005},
        change y base=true,
        y SI prefix=milli,
        x unit=Mbit/s,
        change x base=true,
        x SI prefix=mega,
        xmin = 0,
        xmax = 100000000,
        legend style={
            rounded corners,
            legend cell align={left},
            legend columns=1, 
        },
    ]
    \addplot [thick, dashed, no markers, CoreDarkGray] table [x=idle_slope_swc, y=can_l, col sep=comma] {motivation_20240806_113635_numCanServices_1.csv};
    \addplot [thick, no markers, CoreRed] table [x=idle_slope_swc, y=can_l, col sep=comma] {motivation_20240806_113635_numCanServices_2.csv};
    \addlegendimage{thick, dotted, CoreBlack}
    \legend{
            F1 (F2 inactive),
            F1 (F2 active),
        }
    \end{axis}
\end{tikzpicture}%

%% file: 3_signaling.tex

\section{Negotiating Dynamic Quality-of-Service}
\label{sec:signaling}
\begin{figure}%
    \centering%
    \hspace{0.13\linewidth}
    \includegraphics[width=0.65\linewidth,trim={22pt 22pt 20pt 24pt}, clip=true]{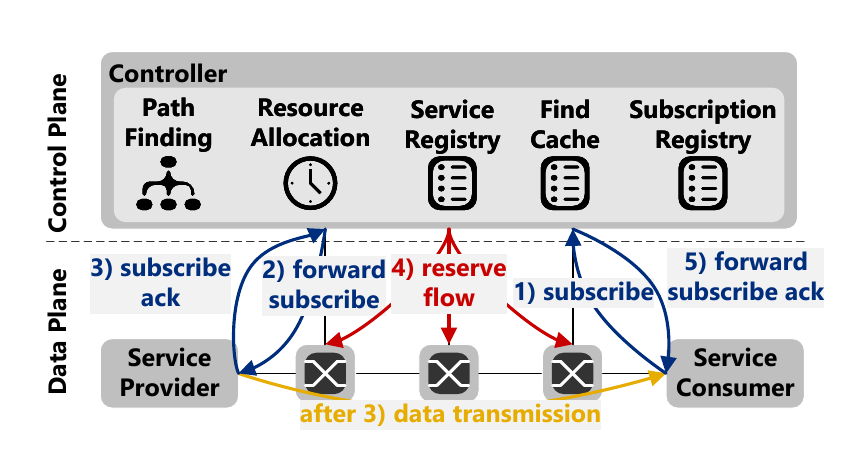}
    \includegraphics[width=0.19\linewidth]{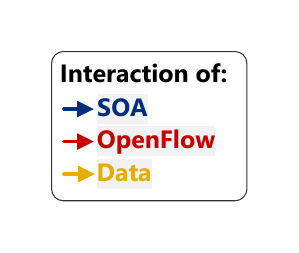}
    \caption{\textcolor{CoreBlue}{SOA} subscription with integrated \ac{QoS} signaling to the \ac{TSSDN} control plane using \textcolor{CoreRed}{OpenFlow} to adapt the network.
    Subsequent \textcolor{CoreYellow}{data transmission} can start after the SOA subscription is established.
    }
    \label{fig:architecture_sdn}
\end{figure}

We first address the identified \ac{QoS} signaling challenges (\cf Sec.~\ref{sec:problem:soa_tsn}) in achieving deadline-compliant service operation within a dynamic SOA: 
\begin{description}
    \item [\textbf{Solution for \problem{1}}:] Providers and consumers embed \ac{QoS} requirements and traffic characteristics directly into the SOA protocol (Sec.~\ref{sec:characteristics}). The controller intercepts this information using OpenFlow (Sec.~\ref{sec:negotiation}).
    This ties the layer 2 \ac{QoS} directly to the layer 5 SOA subscription and fills the gap of a signaling mechanism between the TSN data and control plane. 
    \item [\textbf{Solution for \problem{2}}:] The SOA protocol retains full control over service access (Sec.~\ref{sec:negotiation}), allowing the publisher to reject subscriptions. 
    The controller only acts after receiving the subscription acknowledgment, ensuring proper negotiation before resource allocation.
    \item [\textbf{Solution for \problem{3}}:] The controller determines the required idle slopes to guarantee all deadlines for active flows using the acquired information. 
    Our scheme integrates with different resource allocation and worst-case analysis algorithms, with details provided in Sec.~\ref{sec:worst_case_latency}.
\end{description}

Fig.~\ref{fig:architecture_sdn} illustrates our \ac{TSSDN} subscription process, which integrates these solutions.
Unlike the two-stage procedure (\cf Fig.~\ref{fig:architecture_tsn}), QoS information is embedded directly within SOA messages.
All SOA messages pass through the controller, which caches active services, requests, subscriptions, and their requirements before forwarding them to the respective parties.
For simplicity, the \textit{discovery phase} that takes place beforehand is not shown but follows the same process.
Once a publisher acknowledges a subscription, the controller determines the path, calculates the required idle slopes (\cf Sec.~\ref{sec:worst_case_latency}), and configures all switches along the route.
After setup, switches forward data packets directly to the destination without controller involvement.

Sec.~\ref{sec:deployment} discusses the implications of our approach for automotive deployment.
We illustrate our approach using \ac{CBS} as a proof of concept. 
Nevertheless, the proposed signaling scheme is designed to support other TSN mechanisms as well, such as \ac{ATS}~\cite{ieee8021q-22}.

\subsection{Service Characteristics and Requirements}
\label{sec:characteristics} 
\begin{table}
    \centering
    \caption{QoS options for determining bandwidth and latency bounds.}
    \label{tab:qos_options}
    \begin{tabularx}{\linewidth}{l l L}
    \toprule
    \textbf{Name} & \textbf{Value} & \textbf{Information provider} \\
    \midrule
    \textit{max payload} & Maximum payload per message~[\si{\byte}] & Publisher via advertisement \\
    \textit{min interval} & Minimum interval between messages~[\si{\micro\second}] & Publisher via advertisement \\
    \textit{max burst} & Maximum burst size per interval~[\si{\byte}] & Publisher via advertisement \\
    \textit{deadline} & Maximum \ac{E2E} latency~[\si{\micro\second}] & Subscriber via subscription or publisher via advertisement \\
    \textit{priority} & 802.1Q \ac{PCP} (0-7) & Publisher via advertisement or determined by controller \\
    \bottomrule
    \end{tabularx}
\end{table}

Configuring \ac{TSN} for real-time services relies on a core set of parameters, independent of the specific traffic-shaping method used.
Table~\ref{tab:qos_options} lists the proposed key service characteristics and requirements, derived from \ac{SRP} and \ac{RAP}. 
When allocating resources, the system accounts for publisher properties such as maximum data size, minimum sending interval, and maximum burst behavior.
During subscription, the controller identifies the communication protocols (\eg UDP or TCP) to determine the largest possible frame size.

Publishers and subscribers can specify \ac{E2E} latency deadlines based on application needs or data validity periods, which the system verifies when assigning resources. 
The traffic priority (\ac{PCP}), which affects queueing and shaping in switches, can be predefined by the publisher or dynamically assigned by the controller based on network conditions. 

Our approach is applicable to various \ac{SOA} protocols, provided they support attaching \ac{QoS} information to service discovery and subscription messages. 
Devices that do not support these extensions will ignore them, ensuring backward compatibility with existing implementations.
The primary \ac{AUTOSAR} candidates for automotive \ac{SOA} are \ac{SOME/IP}\cite{a-spsjr-24} and \ac{DDS}\cite{o-dsjr-22}. 
While \ac{DDS} supports \ac{QoS} configuration, it currently lacks link-layer bandwidth reservation capabilities, though it could be extended. 
We focus on \ac{SOME/IP} as it offers a simpler, automotive-specific communication model and is widely adopted in the industry~\cite{iks-aiejr-22}.

To integrate \ac{QoS} signaling, we extend \ac{SOME/IP} native \codeword{offer} and \codeword{subscribe} messages, which already include service descriptions and connection settings.
The protocol defines an extendable configuration option using a list of key-value pairs, which we leverage to embed the \ac{QoS} information seamlessly.

By integrating \ac{QoS} signaling into the subscription process, we eliminate the need for separate layer 2 stream reservations, addressing problem \problem{1} on the data plane level. 
Future extensions could incorporate parameters for redundancy and security~\cite{sms-pdcta-25}, which are beyond the scope of this work.

\begin{figure}
    \centering
    \includegraphics[width=0.7\linewidth,trim={34pt 188pt 46pt 30pt}, clip=true]{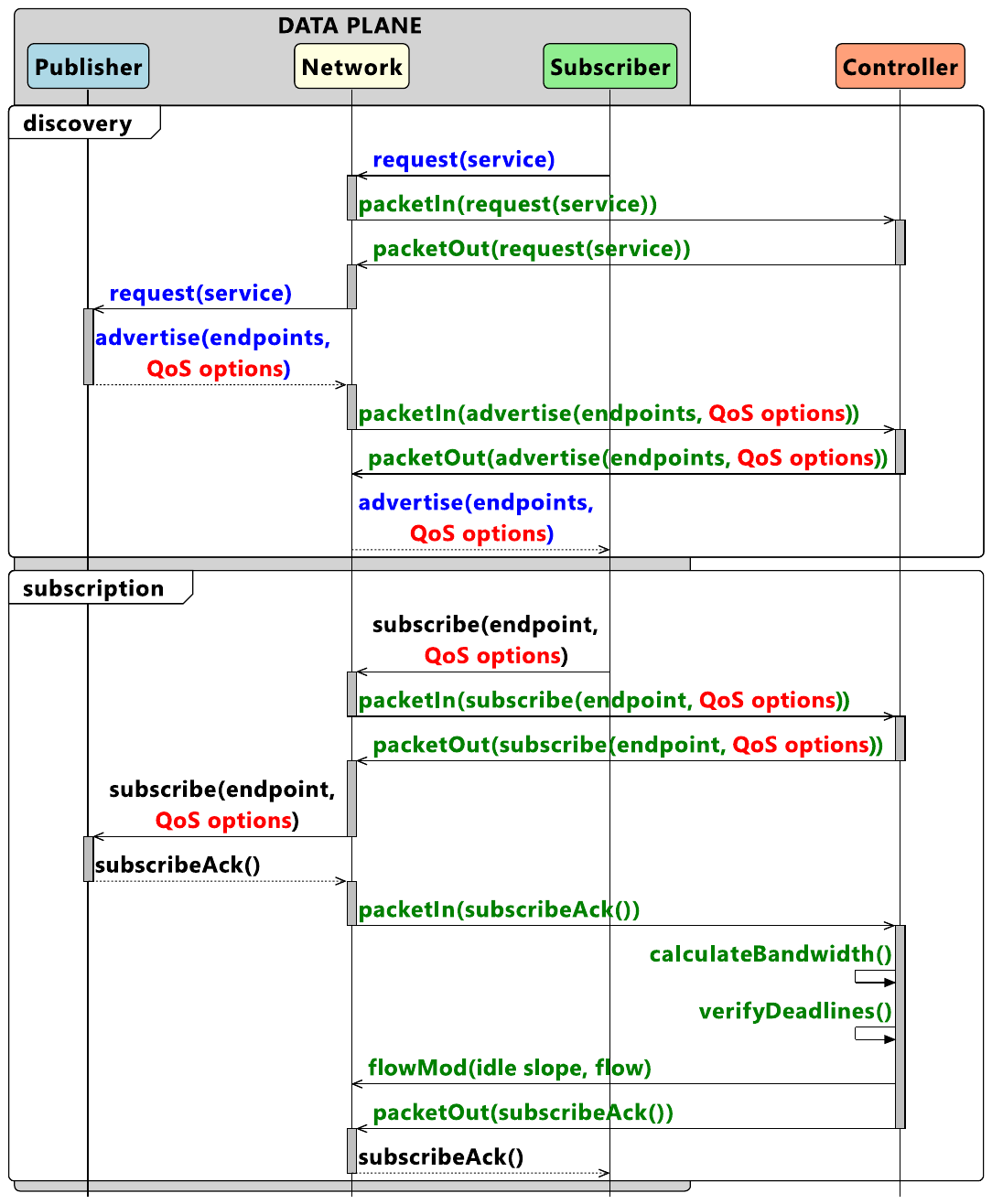}
    \caption{QoS negotiation sequence for \ac{TSSDN} within automotive service discovery that integrates bandwidth allocation and deadline validation. \textcolor{blue}{Multicast messages}, \textcolor{CoreGreen}{control plane operations}, and \textcolor{red}{attached information} are highlighted.
    }
    \label{fig:operations}
\end{figure}

\subsection{Integrated Service Negotiation in TSSDN}
\label{sec:negotiation}
\ac{TSSDN} separates the control plane from the data plane for centralized admission control~\cite{hmks-snsti-19} (\cf Fig.~\ref{fig:architecture_sdn}).
Intercepting network control protocols is a common strategy for optimizing networking objectives in SDN~\cite{krvra-sdncs-15}. 
The central controller serves as a rendezvous point for service discovery, equidistant to all nodes~\cite{hmmsk-dssin-23}, which reduces the signaling overhead between switches compared to a distributed approach.

Fig.~\ref{fig:operations} illustrates how our service negotiation integrates into the automotive service discovery sequence.
Publishers announce services either on request or at regular intervals.
The controller records endpoint locations and replies directly to service requests or queries the network if the endpoint is unknown. 
Subscription requests are registered and forwarded to the publisher.
Services embed \ac{QoS} information into standard service advertisements and subscriptions without altering the existing message sequence.

The controller collects the embedded \ac{QoS} options.
As TSN lacks a standardized signaling protocol between switches and the controller, we use OpenFlow~\cite{onfts025-15} for this purpose, which has been previously applied to \ac{TSSDN}~\cite{hmks-snsti-19,gkbh-sdfjr-19,hmks-stsnv-23}.
The switches relay the publish-subscribe protocol to the controller using \codeword{packetIn} messages, which the controller processes and forwards in the network using \codeword{packetOut} messages.
The controller configures network paths and idle slopes using \codeword{flowMod} messages (details can be found in~\cite{hmks-snsti-19}).
This solves problem \problem{1} on the control plane, as the controller now extracts and utilizes the \ac{QoS} options from the SOA protocol. 

Once a publisher acknowledges a subscription, the controller selects a path for the new flow and calculates new idle slopes to ensure all active flows meet their deadlines.
If validation fails, the controller rejects the subscription, cancels it at the publisher, and notifies the subscriber using the existing means of the service discovery protocol.
On success, the controller updates network devices to enable direct data forwarding and sends the subscription acknowledgment to the subscriber.
The same procedure applies when subscribers unsubscribe or publishers withdraw services.

Because the controller acts only after the acknowledgment of the subscription, the publishers retain the ability to reject subscriptions to enforce access control, providing our solution to \problem{2}.
However, as in previous \ac{SRP}-based approaches (Figs.\ref{fig:architecture_tsn_decentral} and\ref{fig:architecture_tsn_central}), publishers may start the data transfer before the controller installs the flow.
Here, the TSSDN combination has a distinct advantage, as switches only forward traffic that matches a preconfigured flow entry, automatically discarding packets sent before the flow is installed~\cite{hmks-stsnv-23}.
This behavior aligns with automotive \ac{SOA} expectations, where initial packet loss is acceptable when joining a multicast group.

Future work could explore notifying the publisher when flow reservation is complete or installing flows with the subscribe message, although the latter would prevent the publisher from rejecting subscribers. 
Both approaches would alter the service discovery procedure and are beyond the scope of this work.

After negotiation, the controller maintains a complete view of all active subscriptions, service properties, \ac{QoS} requirements, and network topology.
This global awareness enables the controller to compute precise idle slope settings for every switch port and queue, ensuring all active flows meet their deadlines.
By enabling advanced, interchangeable resource reservation schemes for \ac{IVN} use-cases on the control plane, we solve problem \problem{3}.
Sec.~\ref{sec:worst_case_latency} discusses the mathematical models used for these calculations.

\subsection{Automotive Deployment Considerations} 
\label{sec:deployment}
Dynamic services introduce unpredictability into the \ac{IVN}. 
\ac{TSSDN} addresses this by blocking unconfigured services at the network entry point, allowing data transmission only after proper configuration is complete. 
This ensures real-time communication remains unaffected by unauthorized traffic~\cite{hmks-stsnv-23}.

Future research could combine our approach with \ac{TSN} frame replication to improve reliability without sacrificing timing predictability.

Our approach aligns with automotive real-time communication practices, where systems use "fire-and-forget" messaging --- periodic signals sent without retransmission, as outdated messages become obsolete. 
Retransmissions would disrupt traffic patterns and violate reserved bandwidth, making them incompatible with strict timing guarantees. 
For this reason, we treat reliability (guaranteed delivery) and real-time performance as distinct \ac{QoS} choices.
Future research could combine our approach with \ac{TSN} frame replication to improve reliability without sacrificing timing predictability.
Also, we focus on publish-subscribe communication, as one-time request-response exchanges are poorly suited for resource reservation.

While a centralized controller creates a potential single point of failure, redundancy is rarely used in vehicles due to cost, weight, and the fact that vehicles only operate when all critical systems are functional.
Safety-critical functions (\eg braking) typically rely on physically independent backup systems (\eg electric and hydraulic). 
Nonetheless, \ac{SDN} controllers can be logically centralized yet physically distributed across multiple nodes for improved fault tolerance~\cite{krvra-sdncs-15}, though this is beyond the scope of this work.

Service discovery and resource allocation are essential during startup but are not time-critical. 
Vehicles only enter full operation after all services are configured, so immediate real-time communication is not required. 
Still, startup times should remain below \SI{200}{\milli\second} for a seamless user experience~\cite{chrmk-qaasc-19}. 
Centralized SDN-based discovery may introduce delays compared to distributed discovery~\cite{hmmsk-dssin-23}, but protocols already account for potential delays, as both service discovery and \ac{SRP} rely on best-effort communication.
Once flows and reservations are established, subsequent data transfer meets real-time constraints and bypasses the controller. 

In multi-hop topologies, the controller acts as a central rendezvous point equidistant to all nodes, which reduces signaling overhead (\cf Sec.~\ref{sec:eval}). 
Future optimizations could limit network reconfiguration to specific moments, such as service updates or while the vehicle is charging, rather than during operation.

%% file: 5_networkcalculus.tex
\section{Determining Idle Slopes and Latency Bounds}
\label{sec:worst_case_latency}

This section shows how we utilize the central control plane knowledge acquired from our signaling scheme to determine reservations and validate deadlines for dynamic services, addressing the identified challenges in dynamic reservation (\cf Sec.~\ref{sec:problem:dynamic_reservation}): 
\begin{description}
    \item [\textbf{Solution for \problem{3}}:] Advanced centralized resource reservation schemes allow to determine idle slopes that guarantee worst-case latencies for dynamic services. 
	Our scheme integrates with different resource allocation and worst-case analysis algorithms, which can be exchanged, compared, and evaluated for \ac{IVN} use-cases. 
\end{description}

Traditionally, static analysis, \eg \ac{DNC}~\cite{LUDBFF}, was used to determine the worst-case latency for a fixed network configuration and a given set of flows, whereas current dynamic approaches allow for changes in topology and flows, requiring the controller to adapt the network configuration accordingly.
In static network configurations, it is sufficient to verify that the combined local (per hop) latency values along the path of current subscriptions meet the deadline of all flows.
But in a dynamic automotive \ac{SOA}, new subscriptions continuously alter local delays, while already established guarantees must remain valid, necessitating reservation independent bounds. 

Centralized CBS resource reservation for dynamic traffic with worst-case latency analysis has been previously proposed in two state-of-the-art solutions:
\begin{enumerate}
	\item The \codeword{TSN standard}~\cite{ieee8021q-22} provides formulas to compute \textbf{worst-case latency bounds}, assuming that the CBS \textbf{idle slopes} are \textbf{given} by the traffic load. 
	Thus, idle slopes are first selected and subsequently validated.
	\item The \codeword{delay budget} approach, proposed in our previous work~\cite{maile_journal_2022}, uses the analytical framework of \acf{NC} to \textbf{determine} the \textbf{idle slopes} (potentially larger than the traffic load) and \textbf{worst-case latency bounds} based on per-queue delay budgets. 
	The minimal idle slope is selected that ensures compliance with the specified budget.
\end{enumerate}
Both methods ensure that the idle slopes cover at least the reserved flow rate, and the sum of all queue idle slopes at each port is smaller than a maximum threshold (\eg $<$~link rate) before approving a new subscription.

For completeness, we provide details for \ac{E2E} latency calculations, and both dynamic reservation approaches below, using the notations from Table~\ref{tab:notations}.
All formulas are evaluated for each queue in the network individually. 
To simplify notation, we use $\bullet_q$ for queue-specific variables, $\bullet_{(p)}$ for per-priority variables of a queue at the same hop, and $\bullet_{p}$ for global priority variables.

\subsection{End-to-End Latency}
The \ac{E2E} latency of a flow $f$ accumulates along its path $\Phi_f$, with each traversed queue $q$ adding its delay $d_{q}$, including processing, queueing, propagation, and transmission delays.
The latency requirements of a subscription are met if the worst-case \ac{E2E} delay $D_\text{e2e}$ is smaller than the required flow deadline $\mathcal{D}_f$:
\begin{equation}
	\label{eq:e2e-latency}
	D_\text{e2e} = \sum_{q \text{ in } \Phi_f} d_{q} \le \mathcal{D}_f.
\end{equation}
The maximum frame size includes headers, a safety byte, and the inter-frame gap~\cite{ieee8021q-22}.

For \ac{CBS}, the time spent in the queue depends on the idle slope, which enforces a maximum bandwidth. Since a queue is shared by multiple flows, changing the idle slope of one queue can affect the latency of other flows, even in different priorities and network segments~\cite{maile_decentral_2023}.
We distinguish between the worst-case latency for current subscriptions per queue, denoted as $d_q$, and the upper bound for its maximum allowed delay, considering all possible (future) subscriptions, denoted as $D_q$. Thus, $D_q$ is reservation independent and represents a maximum delay budget per queue, which must not be exceeded ($d_q \le D_q$).

\subsection{TSN Standard Approach}
\label{sec:tsn_standard}
TSN standards show multiple formulas for \ac{CBS} worst-case analysis~\cite{ieee8021q-22,BA}, and even reference additional plenary discussions with different formulas.
We apply definitions from~\cite[Annex~L]{ieee8021q-22} (\codeword{Q-WC}) as the most current version in the standard document, to show the implications and differences of the standard values with our previous work. 
We have compared all formulas in the standard in~\cite{maile_decentral_2023}. 

\textbf{Idle Slope: }
The idle slope $\idSL$ is increased with every flow $f$ passing the queue by the flow’s bandwidth which is measured as data transmitted in an observation window.

The \codeword{\ac{CMI}} approach~\cite{ieee8021q-22} uses a pre-defined $\CMI_p$ per priority $p$ as observation window (\eg \SI{125}{\micro\second} for the highest priority~\cite{BA}). 
This results in 
\begin{equation}
	\label{eq:idle_slope_cmi}
	\idSL_{q} = \sum_{\forall f \text{~at~} q} \frac{\MFS_f \cdot \MIF_f}{\CMI_p},
\end{equation}
where $\MFS_f$ is the maximum frame size and $\MIF_f$ the number of frames within $\CMI_p$.

Another approach uses the \codeword{flow interval}, calculating idle slopes based on each flow's individual sending interval
$\FSI_f$ as observation window, which means
\begin{equation}
	\label{eq:idle_slope_stream_interval}
	\idSL_{q} = \sum_{\forall f \text{~at~} q} \frac{\MFS_f}{\FSI_f}.
\end{equation}
A detailed analysis of the implications of using the $\FSI_f$ instead of $\CMI_p$ follows in Sec.~\ref{sec:eval}. 

\input{tab2_notations}

\textbf{Worst-Case Per-Hop Delay: }
The \codeword{Q-WC} procedure assumes to gain an upper bound for the queue delay by calculating a maximum interference due to the CBS shaping. 
Thus, they assume that $d_q$ has a maximum which can be used reservation independent as it will never be surpassed by design as follows.

The maximum value for delay $d_q$ ---called maximum interference delay in the standard--- experienced by a frame in a queue is the sum of queueing, fan-in, and permanent buffer delays ($d_{\mathit{queueing}}$, $d_{\mathit{fan-in}}$, and $d_{\mathit{perm}}$ respectively).
\begin{equation}
	\label{eq:standard-delay}
	d_q = d_{\mathit{queueing}} + d_{\mathit{fan-in}} + d_{\mathit{perm}}
\end{equation}
The queueing delay $d_{\mathit{queueing}}$ is the time it takes to transmit one packet with maximum size $L_{\mathit{max}}$ and all higher-priority packets, defined by:
\begin{equation}
	d_\mathit{queueing} = \begin{cases}
		\frac{L_{max}}{C} & \text{for prio. 7} \\
		\frac{L_{max} + L_{(7)}}{C - \idSL_{(7)}} & \text{for prio. 6}
	\end{cases}
\end{equation}
Other priorities have not been defined.

The fan-in delay $d_{\mathit{fan-in}}$ 
is described as \qq{delay caused by other frames in the same class as frame X that arrive at more-or-less the same time from different input ports}~\cite[p.~2102]{ieee8021q-22}. 
Since the packets of a fan-in burst reside in the buffers when all output bandwidth is used, they cause further delays until they leave the system, reflected by the permanent buffer delay $d_{\mathit{perm}}$. 
$d_{\mathit{perm}}$ is defined to be equal to $d_{\mathit{fan-in}}$.

The formulas do not consider the topology or paths of interfering flows. 
Thus, the delays of \codeword{Q-WC} have to be used carefully, as it has been shown that the delay value for $d_q$ can be unlimited with changing topology~\cite{b-adnjr-17,
maile_decentral_2023}.
Additionally, the formulas are not checked against the actual worst case, which we will elaborate on in Sec.~\ref{sec:eval}.

\subsection{Delay Budget Approach}
\label{sec:delay_budget}
The \codeword{delay budget} method, as proposed in~\cite{maile_decentral_2023,maile_journal_2022,zhao_minimum_2024}, minimizes the idle slopes while considering the delay requirements at each hop. 
Unlike the \codeword{TSN standard}, it determines idle slopes based on flow deadlines, not just traffic characteristics. 
Instead of assuming that \( d_q \) is limited by design, \codeword{delay budget} defines upper bounds (budgets) \( D_q \) for the queueing delays in advance. 
Then, \( d_q \) is selected to approximate \( D_q \) as closely as possible, \ie $ d_q \rightarrow D_q$, with $d_q \le D_q $, in order to minimize the idle slope. 
In our evaluations, we refer to \( d_q \) as \codeword{DB-WC} \textit{current} and to \( D_q \) as \codeword{DB-WC} \textit{independent}.

\textbf{Idle Slope: }
Smaller idle slope values are advantageous for lower priority traffic but, as described in Sec.~\ref{sec:problem}, they increase the queue delay. Thus, the \codeword{delay budget} method utilizes worst-case delay analysis to minimize idle slopes while considering the delay requirements for each hop:
\begin{equation}
\label{eq:idle_slope_delay_budget}
\begin{split}
	\idSL_{q} = \min\big\{\idSL \mid d_q(\idSL) \le D_q\big\}
\end{split}
\end{equation}
where $d_{q}(\idSL)$ represents the worst-case delay for a given idle slope for the current reservations at queue $q$.

\textbf{Worst-Case Queueing Delay: }
For the \codeword{delay budget} method, we need to derive $d_{q}(\idSL)$, for which we use the worst-case analysis called \ac{NC}.
In \ac{NC}, incoming and outgoing traffic is bounded using \textit{maximum arrival} and \textit{minimum service curves}, as illustrated in Fig.~\ref{fig:nc}.
An arrival curve $\alpha(t)$ characterizes the maximum number of bits arriving at a system within any time interval of length $t$~\cite{boudec_network_2012}. 
In our case, the arrival can be determined using the signaled flow specifications. 
The \textit{maximum output} of a queue, denoted as $\alpha^*(t)$, is then used as the worst-case input for the next hop. 
Assuming the simple curves of Fig.~\ref{fig:nc}, $\alpha^*(t)$ can be determined by shifting $\alpha(t)$ by $D_q$. 

A service curve $\beta(t)$ represents the minimum service that a system guarantees to incoming data over any time interval of length $t$~\cite{boudec_network_2012}. Service curves for a network queue are determined through analysis of the worst-case impact of queueing and transmission delays. Depending on the system they also include processing delays, whereas propagation delays are added statically.
In CBS networks, they are in the form of a rate-latency service curve, defined as $\beta_{R,T}(t) = R\cdot[t-T]^+$, where $[x]^+ = \max(0,x)$. 
$R$ is the idle slope of the queue after worst-case interference from other queues, denoted by $T$. 

In short, for each queue in the network, we derive the maximum incoming traffic $\alpha(t)$ and the minimum forwarding behavior $\beta(t)$. 
The worst-case delay $d_q$ for given $\alpha$ and $\beta$ can then be derived as the maximum horizontal distance
\begin{equation}
	d_q(\alpha,\beta) \coloneqq \sup_{t \ge 0} \left\{ \inf \left\{ d \in \mathbb{R}^{+} \;\middle|\; \beta(t+d) \ge \alpha(t) \right\} \right\}.
\end{equation}

\begin{figure}
\centering
	\input{fig6_networkcalculus.tex}
\caption{Basic principles of network calculus to determine the required idle slope $R$ after worst-case interference $T$ to meet the delay budget $D_q$.}
\label{fig:nc}
\end{figure}

For CBS, we can simplify this formula into: 
\begin{equation}
\label{eq:delay_delay_budget}
\begin{split}
	d_q(\idSL) = \frac{\sum_{\{\forall f \text{~at~} q\}} (b_f+r_f\cdot \sum_{\forall q \in \Phi_{q,f}} D_q)}{\idSL} + T_q
\end{split}
\end{equation}
where the numerator is the shifted arrival curve $\alpha^*$ and the denominator with $T_q$ are from the service curve $\beta$, with
\begin{equation}
T_q = \begin{cases}
	\frac{L_{max}}{C} & \text{for prio. 7} \\
	\frac{L_{max}+L_{(7)}}{C} + \frac{\idSL_{(7)}}{C-\idSL_{(7)}}\frac{L_{max}}{C} & \text{for prio. 6}.
\end{cases}
\end{equation}
This can also be graphically seen in Fig.~\ref{fig:nc}. 
As a minimum, the idle slope must match the long-term arrival of the traffic: 
\begin{equation}
	\idSL_q \le \sum_{\forall f \text{~at~} q} r_{f}.
\end{equation}
With these definitions, we can solve Eq.~\eqref{eq:idle_slope_delay_budget} to determine the minimum idle slope.

Note that, to improve readability, we presented only the fundamental concepts of \ac{NC}, using basic functions for illustration. 
We do not cover the effect of traffic shaping, due to link serialization and bandwidth limitations, at this point. 
Traffic shaping improves the value of \( d_q \), reducing the resulting idle slope. 
The definition and application can be directly seen in~\cite[Eq.~(14),~(18)]{maile_decentral_2023}, with the practical implementation available in our open-source tool \ac{DYRECTsn}~\cite{dyrectsn}.

The delay budgets $D_q$ are chosen depending on the flows' deadline and the topology. 
Their choice also influences the success rate of new subscriptions. 
Therefore, \ac{DYRECTsn} implements a meta-heuristic optimization to determine $D_q$ automatically.
Details can be found in~\cite{dyrectsn}.

The \codeword{delay budget} approach can result in sub-optimal configurations if queue delay budgets are poorly allocated. 
However, it is the only method that provides delay guarantees during dynamic reservation as the delay budgets allow to determine a reservation independent upper bound on the latency. 
While these upper bounds may not be as tight as those from static analysis, using \ac{NC} with delay budgets is significantly less computationally expensive and suitable for dynamic networks.

Unlike the \codeword{TSN standard}, the \codeword{delay budget} approach solves problem \problem{3} with proven analytical methods -- the proofs can be found in~\cite{maile_decentral_2023} -- and the calculations consider the traversed paths of all flows. 

%% file: tab2_notations.tex
\begin{table}
    \setlength{\tabcolsep}{15pt}
    \centering
    \caption{Notation for idle slope calculation and latency analysis.}
    \label{tab:notations}
    \begin{tabularx}{\linewidth}{l L}
    \toprule
    \textbf{Variable} & \textbf{Definition} \\
    \midrule
    $C$ & Link capacity\\
    $\Phi_f$ & Path of a flow\\
    $\Phi_{q,f}$ & Path of a flow before queue $q$\\
    $\FSI_f$ & Flow sending interval\\
    $\MFS_f$ & Max. frame size of a flow\\ 
    $\MIF_f$ & Max. number of frames of a flow in an interval\\
    $b_f$ & Max. short term burst of a flow\\
    $r_f$ & Max. long-term sending rate of flow \\
    $\mathcal{D}_f$ & Deadline of a flow\\
    $L_q$ & Max. frame size\\
    $L_{\text{max}}$, $L_\text{min}$ & Max./min. frame size in the network\\
    $d_q$ & Subscription based delay\\
    $D_q$ & Reservation-independent delay\\
    $D_{\text{e2e}}$ & Worst-case \ac{E2E} delay\\
    $\idSL$, $\idSL_q$ & Idle slope (of a queue)\\
    $\CMI_p$ & Class Measurement Interval for priority $p$~\cite[Sec.~L.2]{ieee8021q-22}\\
    \bottomrule
    \end{tabularx}
\end{table}

%% file: fig6_networkcalculus.tex
\begin{tikzpicture}
    \node[anchor=south west,inner sep=0] (image) at (0,0) {\includegraphics[width=.6\columnwidth, trim={20pt 24pt 40pt 25pt}, clip=true]{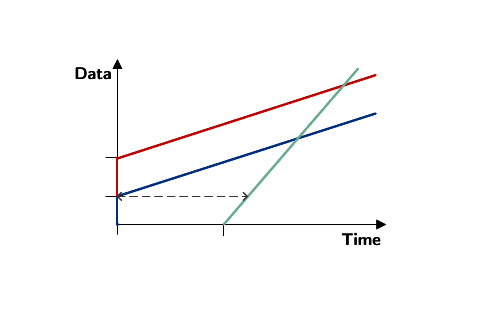}};
    \begin{scope}[{x=(image.south east)},y={(image.north west)}]
        \node[anchor=north]at(0.49,0.12){$T$};
        \node[anchor=east]at(0.18,0.3){$b$};
        \node[anchor=east]at(0.18,0.48){$b + rD_q$}; 
        \node[anchor=east]at(0.4,0.24){$D_q$};
        \node[CoreBlue,anchor=east]at(0.55,0.44){$r$};
        \node[CoreGreen2,anchor=east]at(0.65,0.39){$R$};
        \draw[rounded corners, CoreBlack, thick, fill=white] (0.23,0.64) rectangle (0.64,0.96);
        \draw[CoreBlue, thick] (0.25,0.9) -- (0.29,0.9) node[black,anchor=west,font=\normalsize] {$\alpha_{b,r}(t) = b + rt$};
        \draw[CoreGreen2, thick] (0.25,0.8) -- (0.29,0.8) node[black,anchor=west,font=\normalsize] {$\beta_{R,T}(t) = R\cdot[t-T]^+$};
        \draw[CoreRed, thick] (0.25,0.7) -- (0.29,0.7) node[black,anchor=west,font=\normalsize] {$\alpha^*(t)$};
    \end{scope}
\end{tikzpicture}%

%% file: 6_evaluation.tex
\section{Evaluation}
\label{sec:eval}
We evaluate our service negotiation scheme using both a synthetic parameter study and a realistic \ac{IVN} scenario. 
While previous work has acknowledged the complexity of determining idle slopes in \acp{IVN}~\cite{mvnb-ipcjr-18}, existing bandwidth reservation schemes have not been systematically compared in detail.

Our comprehensive evaluation framework combines analytical worst-case calculations with simulations of realistic network configurations. 
This dual approach makes it possible to systematically generate and analyze edge-case scenarios, which is a key advantage over implementations in real hardware.
The worst-case analysis leverages formal methods (\cf Sec.~\ref{sec:worst_case_latency}) to derive theoretical bounds, while simulations capture the complex interplay of flows in dynamic environments. 
The latter invalidates worst-case assumptions by providing dedicated counterexamples and offers quantitative insights into setup times and bandwidth utilization.
In combination, these methods enable us to quantify the impact of reservation strategies on resource allocation and network delay in practical deployment scenarios. 
By demonstrating the real-world feasibility of our approach to guarantee strict latency limits for dynamic services, we establish its relevance for future automotive designs.

We restrict our analysis to \ac{CBS} due to its suitability for dynamic traffic in cars (\cf Sec.~\ref{sec:background_tsn}). 
For worst-case analysis, we focus on deterministic algorithms based on rigorous mathematical models for deadline validation, as detailed in Sec.~\ref{sec:worst_case_latency}.
Hence, we do not use current \ac{DRL}-based approaches that often focus on \ac{TAS} implementations, and -- to the best of our knowledge -- do not provide idle slopes that guarantee deadlines for all active flows.
We further discuss the comparison to \ac{ML} approaches in Sec.~\ref{subsec:ml_limitations}.
Nevertheless, the proposed framework generalizes to other shapers and centralized configuration algorithms, which ensures broad applicability.

\subsection{Simulation Environment and Configuration}
\label{sec:eval_env}
\begin{figure}
    \centering
    \includegraphics[width=0.6\columnwidth, trim={18pt 18pt 18pt 18pt}, clip=true]{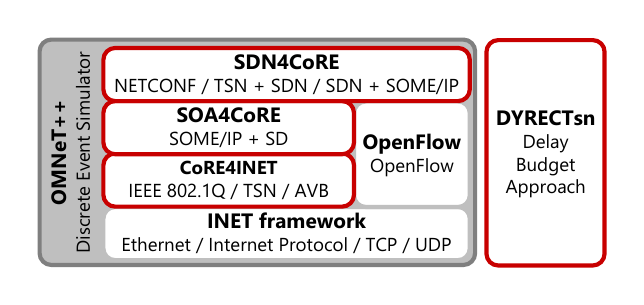}
    \caption{Simulation environment with simulation models for TSN, automotive SOA, OpenFlow, and TSSDN. The \ac{DYRECTsn} Python framework implements the delay budget approach.
    \textcolor{CoreRed}{Highlighted} frameworks are published and maintained by authors of this paper.
    The environment and scenarios are available at \url{https://github.com/CoRE-RG/vehcom25-soa-strict-cbs-latency}.}
    \label{fig:simulation_environment}
\end{figure}
Fig.~\ref{fig:simulation_environment} depicts the simulation environment, which is based on OMNeT++~\cite{omnetpp} and the INET framework~\cite{inet-framework}.
The CoRE4INET~\cite{mkss-smcin-19}, SOA4CoRE~\cite{hmmsk-dssin-23}, OpenFlow~\cite{kj-oeojr-13}, and SDN4CoRE~\cite{hmks-smsdn-19} simulation models implement TSN, automotive SOA (\ac{SOME/IP}), OpenFlow, and TSSDN, respectively. 
The \ac{DYRECTsn} Python framework~\cite{dyrectsn} implements the \codeword{delay budget} approach. 
CoRE4INET, SOA4CoRE, SDN4CoRE, and \ac{DYRECTsn} have been previously published and are maintained by the authors of this paper.
All experiments were conducted on a desktop PC (Windows 11, 64 GB RAM, Intel\textcopyright 13900K).

We add our service negotiation and reservation schemes to SDN4CoRE and implement the \ac{QoS} configuration options for \ac{SOME/IP} in SOA4CoRE. 
Further, we implement the simulated scenarios in \ac{DYRECTsn} to generate a configuration, which we then apply in the simulator. 
We also implement the TSN worst-case analysis of the standard in a Python framework. 
All the frameworks are open-source, and the assembled environment with all changes and scenarios is available on GitHub\footnote{Simulation environment and scenarios available at: \url{https://github.com/CoRE-RG/vehcom25-soa-strict-cbs-latency}}.

With this comprehensive environment, we can evaluate the proposed service negotiation in combination with different reservation schemes and worst-case analysis.
We compare the \codeword{TSN standard} worst-case analysis \codeword{Q-WC} for the two idle slope configurations \codeword{CMI} and \codeword{flow interval} against the \codeword{delay budget} worst-case analysis \codeword{DB-WC} and its chosen configuration.
Graphs use the following additional abbreviations: \codeword{flow interval} (FI), and \codeword{delay budget} (DB).

The frameworks offer a wide range of configuration parameters, which we set to match the setup in~\cite{hmmsk-dssin-23}: 
All switches have a hardware forwarding delay of \SI{8}{\micro\second}. 
The OpenFlow messages processing time in the controller application is \SI{100}{\micro\second}, based on the worst-case performance of the best-performing controller implementation we have evaluated in previous work~\cite{rhmks-rapesc-20}.
We assume the switch processing time is similar and set it to \SI{100}{\micro\second} but could not find any data in the literature on the performance of OpenFlow processing on switches.
The controller and the switches can handle multiple OpenFlow packets in parallel. 

\subsection{Study of Bandwidth Reservation and Worst-Case Analysis}
\label{sec:eval_study}
\begin{figure}
    \centering
    \includegraphics[width=0.7\columnwidth, trim={48pt 48pt 22pt 24pt}, clip=true]{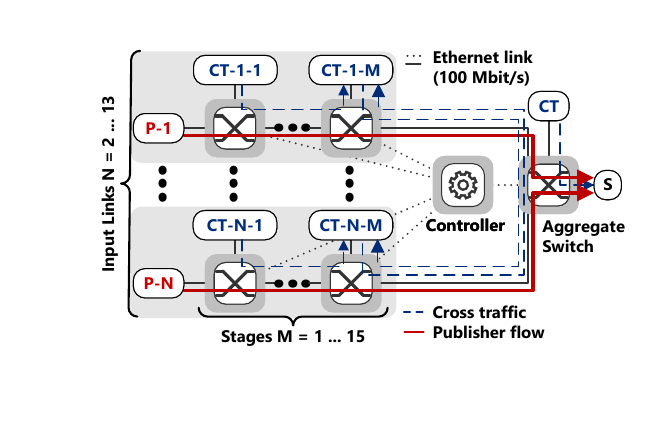}%
    \caption{Study to analyze the influence of cross traffic and network topology on service negotiation, idle slopes, and the analytical worst-case. 
    }
    \label{fig:cbsstudy}
\end{figure}
We compare the reservation schemes in a parameter study shown in Fig.~\ref{fig:cbsstudy}.
Similar studies have been proposed in~\cite{b-adnjr-17, maile_decentral_2023} focusing only on static idle slopes.
We evaluate dynamic service negotiation, bandwidth reservation schemes, and worst-case analysis.

Publishers in two to thirteen input links ($N$) send to a subscriber via a series of one to fifteen switch stages (M), paired with a cross traffic (CT) generator. 
An aggregate switch merges the input links and connects to the subscriber. 
The link bandwidth is \SI{100}{\mega\bit\per\second}.

Publishers send one frame every \SI{125}{\micro\second} with highest priority. 
Their frame size varies with the number of input links to achieve a total of \SI{75}{\mega\bit\per\second} sent to the subscriber, with each publisher transmitting frames of $\SI{1171}{Byte}/N - \SI{12}{Byte}$ to account for inter-frame gaps.

Each CT targets one link in the publisher's path, sending traffic to the next CT node, which allows for maximum interference.
The last node in the chain sends to the next input link via the aggregate switch.
A final CT sends to the subscriber through the aggregate switch.

In our study, the CT along the stages either sends full-size Ethernet frames as best effort (BECT), or frames with the same priority as the publishers (PCT).
The PCT is configured to utilize the remaining bandwidth left by the publisher to reach the total of \SI{75}{\mega\bit\per\second} on each input link, thus the PCT frame size is $\SI{1171}{Byte} - \SI{1171}{Byte}/N - \SI{12}{Byte}$. 
The CT interval is set to \SI{100}{\milli\second} to produce repeatable burst patterns.

\subsubsection{Service Negotiation Delay}
\begin{figure}
    \input{fig9_setuptime_cbsstudy.tex}
    \caption{Time to set up all subscriptions for input links (N) and stages (M). Number of service negotiations are ($N \cdot (M+1) + 1$).}
    \label{fig:study_setuptime}
\end{figure}
All flows from publishers and CT are negotiated and configured using our integrated SOA signaling for TSSDN.
To avoid impacting the rest of the study, traffic generation starts only after all negotiations are complete. 

Fig.~\ref{fig:study_setuptime} shows the total setup time for all subscriptions, from the start of the first negotiation to the completion of the last, across varying numbers of input links and stages.
Setup time increases with the number of connections when increasing the input links but remains around \SI{1}{\milli\second}. 
The number of stages has only a small impact on setup time.
This efficiency is primarily due to the controller acting as a rendezvous point with constant distance to every node, which improves setup time in \ac{SDN} as we have shown in previous work~\cite{hmmsk-dssin-23}.

The simulation is configured for parallel processing at the controller with a constant processing time of \SI{100}{\micro\second} per request including idle slope calculation and worst-case estimation. 
Nonetheless, the communication overhead for negotiation remains low, even for many nodes and services -- \SI{1.2}{\milli\second} for the max. 209 service negotiations ($N \cdot (M+1) + 1$).

\subsubsection{Idle Slope Configuration}

The idle slope is adjusted whenever a new subscription is added. 
Fig.~\ref{fig:idleslopeconfig} shows the idle slope configuration for the stage switches in a 5-stage chain. 
The idle slope along the stages includes the bandwidth required for both publishers and the PCT regardless of whether the PCT is active.

With the \codeword{flow interval} approach, the bandwidth reservation for the PCT is small due to its slow send interval of \SI{100}{\milli\second}.
As the publisher frame size decreases with more input links, the idle slope also decreases. 
Using a \codeword{CMI} of \SI{125}{\micro\second} results in a constant idle slope of \SI{75}{\mega\bit\per\second} for all stages, as the combined publisher and PCT frame sizes add up to the same value for each configuration.
The \codeword{delay budget} approach determines idle slopes along the path that meet the flow deadlines, potentially resulting in different idle slopes for the switch in each stage.
On average, the \codeword{delay budget} idle slope configuration falls between the two standard approaches, reserving about \SI{50}{\mega\bit\per\second}.

At the aggregate switch, the publisher traffic from all input links consistently adds up to \SI{75}{\mega\bit\per\second}. 
This is reflected in the idle slope with the \codeword{flow interval} and \codeword{flow interval} approaches. 
The \codeword{delay budget} approach considers flow deadlines and queue delay budgets, which in our case also results in a \SI{75}{\mega\bit\per\second} idle slope.

\subsubsection{Worst-case Analysis}

The study aims to cause large interference for the publisher flows. 
When delayed by CT, publisher packets can accumulate and cause a quasi-burst~\cite{b-adnjr-17}. 
This means that although the publisher flows do not exceed their reserved bandwidth, they temporarily occupy bandwidth allocated for the PCT, which is unused because of the slow send interval.
The number of consecutive packets in these quasi-bursts increases with the chain length, causing significant delays and queue buildup at the aggregate switch where they merge with other publisher flows. 
This phenomenon is explained in~\cite{b-adnjr-17} and in previous work we demonstrated that it can even occur in a network where all flows have a path length of two~\cite{maile_decentral_2023}.

\begin{figure}
    \input{fig10_idleslopeconfig.tex}%
    \caption{Idle slope at the stage switches for M=5 stages for the CMI, flow interval (FI), and delay budget (DB) approaches. For the delay budget approach, the idle slope varies between the stage switches, showing minimum, maximum, and average.}
    \label{fig:idleslopeconfig}
\end{figure}
\begin{figure}
    \input{fig11_aggdelayq.tex}%
    \caption{Queueing delay at the aggregate switch for four input links under best-effort (BECT) and high-priority (PCT) cross traffic for the CMI and flow interval (FI) approaches comparing empirical results from simulation (Sim) with the analytical TSN worst-case (Q-WC).}
    \label{fig:aggdelayq}
\end{figure}

The \codeword{Q-WC} analysis from~\cite[Annex~L]{ieee8021q-22} claims to provide maximum per-hop queueing delay for a given idle slope configuration.
Fig.~\ref{fig:aggdelayq} shows the delay at the aggregate queue for four input links from simulations with BECT and PCT, and the corresponding results from \codeword{Q-WC} analysis.
The constant worst-case for different numbers of stages does not account for the chain length, so it fails to capture the quasi-burst effect. 
It is evident that our BECT simulations with idle slopes determined for a \codeword{CMI} surpass the predicted \codeword{Q-WC}.
For the \codeword{flow interval} approach, the \codeword{Q-WC} is very conservative compared to our simulations.

\begin{figure}
    \input{fig12_study_e2e_combined.tex}%
    \caption{\ac{E2E} delay for thirteen input links under best-effort (BECT) and high-priority (PCT) cross traffic for the CMI, flow interval (FI), and delay budget (DB) approaches. Empirical values from simulations (Sim) are compared against analytical worst-case (WC) for the TSN standard method (Q-WC) at the top, and the delay budget method (DB-WC) at the bottom. The delay budget method considers the worst case for the current reservation and the maximum possible (independent) reservation.}
    \label{fig:study_e2e_db}\label{fig:study_e2e_q}
\end{figure}

Fig.~\ref{fig:study_e2e_q} shows the maximum \ac{E2E} delay of the publisher flows for thirteen input links and different chain lengths from simulations and worst-case analysis.
The top shows the \codeword{Q-WC} for the \codeword{CMI} and \codeword{flow interval} approaches.
The bottom shows the results for the \codeword{delay budget} approach and the worst-case analysis from \ac{DYRECTsn}, providing delay bounds for the current reservation (\codeword{DB-WC} current) and the reservation independent worst-case that remains valid for future flow updates (\codeword{DB-WC} independent).

All approaches show a similar \ac{E2E} delay in simulation when only BECT is active.
However, the smaller reservation for the \codeword{flow interval} approach increases the potential delay caused by the PCT, as publishers must wait longer for the credit to accumulate. 
Again, the \codeword{Q-WC} does not provide a valid delay bound, this time for the \codeword{flow interval} approach.
The \codeword{delay budget} approach provides a valid configuration that stays below the worst-case analysis from \ac{DYRECTsn}.

With our user-configured queue delay budgets of \SI{300}{\micro\second} for the stages and \SI{5}{\milli\second} for the aggregate switch, \ac{DYRECTsn} was unable to schedule all flows for thirteen input links with more than thirteen stages.
A valid configuration may exist, for a different configuration.
This shows that the \codeword{delay budget} approach does not schedule flows that would violate the configured latency bounds.

\begin{figure}
    \centering
    \includegraphics[width=.95\columnwidth, trim={22pt 56pt 12pt 53pt}, clip=true]{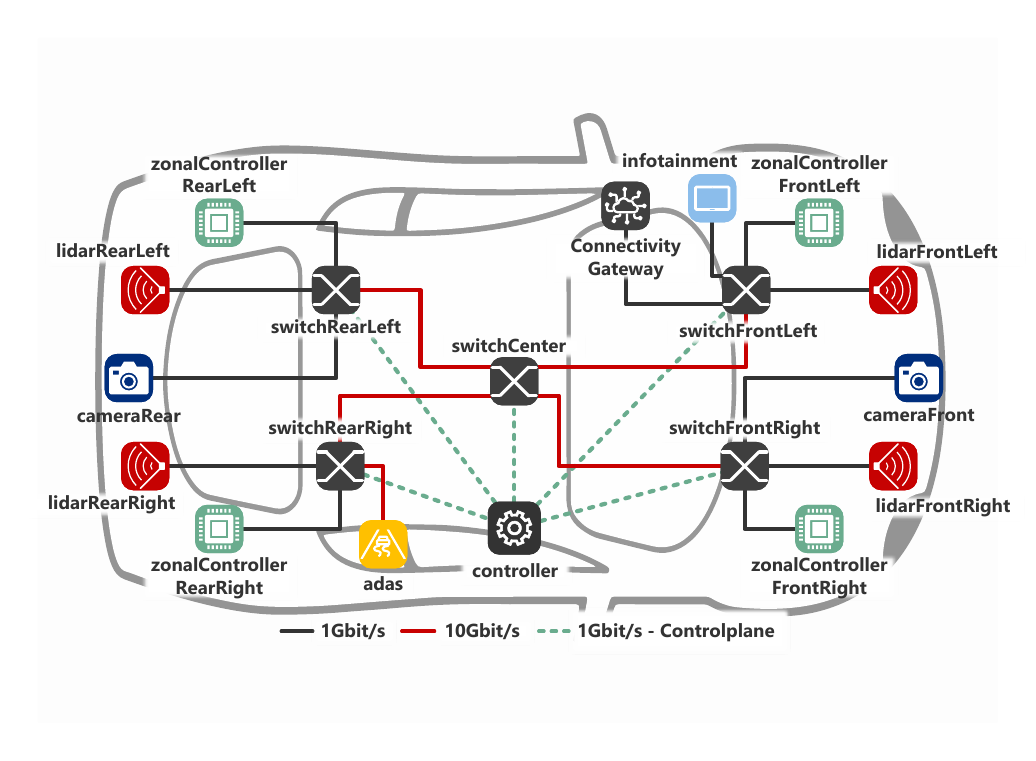}
    \caption{\ac{IVN} topology based on~\cite{mhlks-fsaad-24} modified to a star topology with upgraded \SI{10}{\giga\bit\per\second} links between the switches and a central controller.}
    \label{fig:car_network}
\end{figure}
\subsection{In-Car Network Case Study}
\label{sec:eval_car}
To show the differences of the reservation schemes in practice, we use our realistic open-source \ac{IVN} model previously published in~\cite{mhlks-fsaad-24}.
We convert all traffic sources and sinks into \ac{SOME/IP} services while maintaining the original traffic patterns and communication relations, keeping the transport protocols unchanged. 
The original topology featured a redundant ring backbone, which we simplified to a star topology (Fig.~\ref{fig:car_network}), as redundancy is beyond the scope of this work.
We add a central controller connected to all switches. 

The in-car traffic is organized as follows (details can be found in~\cite{mhlks-fsaad-24}):
Timed control traffic using synchronous gates (PCP 7), which we keep as is,
lidar and video streams using CBS (PCP 6),
embedded CAN signals to which we also apply CBS (PCP 5),
and best effort TCP traffic from the connectivity gateway.
Out of the total 216 streams, we focus on negotiating guaranteed latency for 212 UDP publishers (including LIDAR, camera, and CAN signals) and 448 subscribers, averaging 2.1 subscribers per publisher (min. of 1, max. of 4).
The original \ac{IVN} does not provide deadlines for the anonymized flows, so we set a deadline of \SI{1}{\milli\second} for all flows.

With this traffic setup, only the \codeword{flow interval} approach can provide bandwidth reservations below \SI{1}{\giga\bit\per\second} on any port. 
With the implemented heuristic for determining delay budgets, \ac{DYRECTsn} was not able to schedule all flows, however, a valid configuration may exist.
Following the work of Walrand \etal\cite{wtm-avnjr-21}, we thus combine different link speeds using \SI{1}{\giga\bit\per\second} links connecting the devices to the backbone, and \SI{10}{\giga\bit\per\second} links between the switches as a central infrastructure.
The ADAS node receives all camera and lidar data, also requiring a \SI{10}{\giga\bit\per\second} link to the backbone.

\subsubsection{Service Negotiation Delay}
All services start at \SI{90}{\milli\second} simulation time.  
In contrast to the synthetic study, data transmission starts right after the reservation concurrent to ongoing negotiations.
The negotiation time for all subscriptions is \SI{2.2}{\milli\second}.
However, \ac{SOME/IP} suggests scattering the service start up (commonly set between \SIrange{10}{100}{\milli\second}) to avoid network congestion.
We set a uniform distribution between 0 and \SI{10}{\milli\second} extending the negotiation time to \SI{11.1}{\milli\second} (see Fig.~\ref{fig:car_idleslope}).

We also measure the \ac{DYRECTsn} performance to calculate the idle slopes for each subscription, which is \SI{0.4}{\milli\second} on average (min. \SI{0.07}{\milli\second} and max. \SI{1.3}{\milli\second}). 
While this is on average higher than the \SI{100}{\micro\second} processing time per request set at the controller (\cf Sec.~\ref{sec:eval_env}), the Python-based implementation is not optimized for performance.

\begin{figure}
    \input{fig12_car_idleslopeconfig.tex}%
    \caption{Idle slope configuration during the \ac{IVN} setup at switchCenter towards switchRearRight for the CMI, flow interval (FI), and delay budget (DB) reservation schemes and priorities 5 (CAN), 6 (camera, lidar), and their sum.}
    \label{fig:car_idleslope}
\end{figure}
\subsubsection{Idle Slope Configuration and End-to-End Delay}

The idle slope configuration is updated for each of the 450 subscriptions.
The timed-control traffic using synchronous gates is unaffected by dynamic reservation, as it is not negotiated.
For the \codeword{CMI} and \codeword{flow interval} approaches, we reduce the available bandwidth at each port by the time the CBS gates are closed. 
However, the formulas do not consider gate closing times, resulting in unchanged idle slopes.
In contrast, the \codeword{delay budget} idle slopes guarantee delays by including gate closing times in their model. 
The simple gate control list enables efficient calculation by considering reduced rate and increased initial latency for every port's service, thereby lower bounding the approach of~\cite{zhao_improving_2021}.

Fig.~\ref{fig:car_idleslope} shows the idle slope configuration over time at switchCenter towards switchRearRight for the reservation schemes.
The setup duration is \SI{11.1}{\milli\second}, starting at \SI{90}{\milli\second} simulation time, with idle slope increasing with each new subscription.
The reservation schemes show significant differences in idle slope configuration for PCP 5 CAN signals, PCP 6 lidar and camera streams, and the combined port reservation.
The embedded CAN communication generates small packets with long intervals (\SIrange{10}{1000}{\milli\second}) causing a tradeoff between idle slope and maximum delay (\cf Sec~\ref{sec:problem}-\ref{sec:problem:dynamic_reservation}).
The combined traffic of all cameras and lidars is about \SI{800}{\mega\bit\per\second}, which may require an even larger idle slope to guarantee a \SI{1}{\milli\second} deadline.

Table~\ref{tab:car_combined_e2e_is} shows the port idle slope configurations and the service \ac{E2E} latencies. 
The CAN signals have a very low reservation of about \SI{1}{\mega\bit\per\second} with the \codeword{flow interval} approach.
The \codeword{CMI} and \codeword{delay budget} set a significantly higher idle slope for these signals with a maximum of \SI{398}{\mega\bit\per\second} and \SI{512}{\mega\bit\per\second}, respectively.
In turn, the \codeword{flow interval} approach is not able to meet the required deadline with a maximum delay of \SI{148}{\milli\second} in simulation.

The camera and lidar streams have a short interval (\SIrange{65}{150}{\micro\second}), thus the difference between \codeword{flow interval} and \codeword{CMI} is smaller.
The \codeword{delay budget} approach reserves a much higher idle slope with a maximum of \SI{2.2}{\giga\bit\per\second}, due to the per port delay budgets of only about \SI{200}{\micro\second} determined by the heuristic in \ac{DYRECTsn}.

The \codeword{CMI} approach seems to provide a valid configuration, however, as shown before there is currently no solution to verify deadlines at runtime.
The simulation might not reflect the worst case as we do not consider any best effort traffic or maximize service interference.
The independent guarantee of the \codeword{delay budget} approach for all services are between \SI{408}{\micro\second} and \SI{824}{\micro\second} fulfilling the set latency requirements and accounting for possible updates.

\begin{table}
    \centering
    \setlength{\tabcolsep}{3pt}
    \caption{Switch idle slopes and \ac{E2E} latencies for all services of a priority from simulations between the reservation schemes. 
    Latencies surpassing the set deadline of \SI{1}{\milli\second} are \textcolor{red}{highlighted}.}
    \label{tab:car_combined_e2e_is}
    \input{tab3_car_combined_is_e2e.tex}
\end{table}

\subsection{Key Findings}
\label{sec:eval_discussion}

\begin{description}
    \item[Findings regarding \problem{1}:] The proposed signaling scheme bridges the layer 5--2 \ac{QoS} gap, by embedding reservations directly in \ac{SOA} messages.
\end{description}

Service negotiation completed in \SI{2}{\milli\second} for 450 subscriptions in the \ac{IVN}, in line with the synthetic study (\SI{1.2}{\milli\second}).
With \SI{10}{\milli\second} \ac{SOME/IP} scatter, negotiation time slightly extended to \SI{11.1}{\milli\second} --  well below the \SI{200}{\milli\second} user-experience threshold~\cite{chrmk-qaasc-19}.  
At the same time, scattering reduces controller requests rates, which can be beneficial for the overall network performance.
Once flows and reservations are established, real-time data transfer bypasses the controller.

In contrast to the two-stage procedure (\cf Sec.~\ref{sec:problem:soa_tsn}), our scheme preserves lower-layer socket abstractions and OS compatibility.
The controller autonomously derives low-level parameters without stack modification.
Error handling follows standard \ac{SOA} procedures.
Recovery procedures for message loss and timeouts can remain the same, however they need to be set to account for potential delays due to the controller interactions.
A path is only installed, when a subscription is acknowledged and the reservation succeeds preventing propagation of unreserved traffic. 
Failed validations trigger automatic subscription cancellation through existing \ac{SOA} means.
Our approach simplifies handling of timeouts and ungraceful cancellations, improving resource efficiency.

\begin{description}
    \item[Findings regarding \problem{2}:] Access control becomes \ac{SOA}-centric, with the controller enforcing publisher decisions through centralized management. 
\end{description}

The controller maintains the TSN multicast group access, restricting who can join to participants dictated by the \ac{SOA}.
This architecture supports additional authentication layers while maintaining strict adherence to subscription policies.

\begin{description}
    \item[Findings regarding \problem{3}:] Centralized reservation using the \codeword{delay budget} method guarantees strict latency for dynamic services.
\end{description}

The evaluated reservation schemes demonstrate distinct performance characteristics. 
The \codeword{flow interval} method risks severe misconfigurations as slow send intervals result in insufficient idle slope reservations, causing excessive delays. 
While \codeword{CMI} approach performed better in configuring idle slopes to meet delay constraints, it shares the limitation of unreliable delay prediction~\cite{b-adnjr-17,maile_decentral_2023}. 
Only the \codeword{delay budget} approach consistently maintains hard latency constraints by proactively minimizing per-hop idle slopes to meet specified budgets, rejecting any reservation that would violate latency guarantees. 

The order of operation between the approaches already gives an insight into their efficiency. 
Unlike \codeword{TSN standard} procedures that validate deadlines after allocation, the \codeword{delay budget} approach enforces deadline constraints during the reservation process itself.
In turn, the delay guaranteed idle slopes from the \codeword{delay budget} approach are up to three times higher than the \codeword{flow interval} method (\cf Fig.~\ref{fig:car_idleslope}), which requires significant overprovisioning of network resources.

\section{Discussion}
\label{sec:discussion}
Our study reveals key insights into centralized \ac{CBS} reservation for in-car networks and demonstrates the effectiveness of our signaling approach embedded in the \ac{SOA}.
While our approach addresses the three challenges outlined in Sec.~\ref{sec:problem}, limitations and open questions remain. 

\subsection{What Are the Limitations of Our Approach -- and How Can They Be Mitigated?}
\label{subsec:limitations}
Our approach faces deployment challenges, as already discussed in Sec.~\ref{sec:deployment}, which include start-up delays caused by control-plane signaling, and the controller instance acting as a single point of failure.
In practice, these may cause reservation failures or delays that exceed user-experience thresholds, a common limitation in dynamic reservation systems.
To mitigate impact of such failures and delays, services must verify functionality post-reconfiguration before the vehicle enters operational mode. 
Once reservations and flows are established, real-time data transfer operates independently of the controller, reducing reliance on its continuous availability.
Stable service configurations across multiple system startups suggest an optimization opportunity:
network updates could be restricted to discrete events (\eg software updates or charging sessions),
thereby minimizing the start-up delay and impact of control plane failure.

Another limitation is early transmission, a timing vulnerability which emerges when publishers transmit data before reservation completes, which remains unchanged from the two-stage reservation procedure (\cf Sec.~\ref{sec:problem}).
Our \ac{TSSDN}-based approach mitigates early transmissions using ingress blocking of unknown flows until reservation succeeds, protecting other streams from unplanned cross traffic.
This ensures real-time operation takes precedence over message loss of early data transmissions, preventing unauthorized traffic from disrupting time-critical streams.
While changing the discovery procedure could resolve this (\eg the controller notifying the publisher upon successful reservation), such protocol modifications are deferred to future work.

Reliable data delivery presents a further challenge, as retransmissions conflict with strict resource reservations by disrupting scheduled traffic. 
We treat reliable delivery and real-time guarantees as orthogonal \ac{QoS} requirements, prioritizing the latter in this work. 
Future extensions could integrate redundancy mechanisms, such as multi-path bandwidth reservation via \ac{TSN} path redundancy (IEEE 802.1CB -- FRER), to enhance reliability without sacrificing determinism.

Lastly, while our approach retains the access-control mechanisms of distributed \ac{SOA} architectures, the role of the central controller in network configuration introduces a potential access-control risk. 
This risk is not unique to our design as all network decision-making systems influence traffic routing and thus require safeguards.
This underscores the critical need to secure the controller against unauthorized access or manipulation, as its decisions ultimately dictate traffic routing policies.

\subsection{Which Resource Reservation Schemes Suit IVNs?}
\label{subsec:reservation_schemes}

Our study confirmed -- consistent with prior findings~\cite{b-adnjr-17,maile_decentral_2023} -- that estimates derived from the \codeword{TSN standard} approach fail to provide reliable delay bounds, rendering it inadequate for \acp{IVN}. 

While static analysis tools (\eg \ac{DNC}~\cite{LUDBFF}) could theoretically validate \codeword{TSN standard} reservations, full network analyses remain computationally expensive and often take hours on current CPUs. 
This precludes deployment in dynamic, safety-critical \acp{IVN}.

By comparison, \ac{DYRECTsn} computed idle slopes for \ac{IVN} subscriptions in \SI{0.4}{\milli\second} on average, which makes it applicable in live systems.
Further optimizations may reduce this runtime.
For now, the \codeword{delay budget} method remains the only reliable solution for dynamic reservations with limited latency.

\subsection{Can Machine Learning Do Better?}
\label{subsec:ml_limitations}
\ac{ML} approaches promise to simplify, automate, and optimize the configuration of network parameters~\cite{kkw-mlcjr-22} but require extensive, domain-specific training data, which are often difficult to obtain and generalize for \ac{IVN} in practice~\cite{mhlks-fsaad-24}. 
In contrast, our approach proposes an efficient integration of \ac{QoS} signaling for existing automotive \ac{SOA} protocols. 
This method communicates known traffic characteristics and \ac{QoS} requirements to the network controller and hence neither requires extensive training, nor introduces delays from learning characteristics prior to scheduling flows.

\ac{DRL}-based methods for traffic classification, flow-aware scheduling, and routing~\cite{hzdxy-defjr-23,cl-dnsjr-23,cyzz-rrsjr-25}
demonstrate improvements in flow scheduling, average latency, and load balancing.
However, they fail to provide strict latency guarantees for established flows, which are critical for \acp{IVN}.
Related evaluations showed improvements in average latency but could not demonstrate real-time capability in \acp{IVN}, in which the worst-case maximum latency is critical. 
For example, prior work achieved approximately 95\% schedulability for up to 200 flows~\cite{hzdxy-defjr-23} in fixed time-slots,
whereas our \ac{IVN} use case demands deterministic deadlines for 450+ subscriptions.
In practice, our approach enables a fast and reliable method to determine idle slopes and worst-case latencies before subscriptions are established.
This ensures that the network can meet the required deadlines. 
To be applicable in safety critical applications, ML mechanisms must provide explainable and strict delay guarantees for selected idle slopes.

Overall, \acp{IVN} require strict latency guarantees, which existing analytical models, such as the proposed \codeword{delay budget} approach, can provide.
Crucially, our approach offers fast and reliable control plane signaling with deadline awareness, which is a prerequisite for safety-critical systems that cannot be inferred from traffic patterns alone. 
As such, analytical models and signaling protocols provide a rigorous solution that outperforms the proposed \ac{ML} and \ac{DRL} concepts for safety-critical applications.

\subsection{Are Compatibility and Complexity Affected?}
\label{subsec:compatibility_complexity}
Our signaling scheme bridges the \ac{QoS} gap between layer 5 \ac{SOA} and layer 2 \ac{TSN} to provide latency guarantees in dynamic automotive environments. 
Although it is not backward-compatible with legacy \ac{TSN} devices, it strengthens interoperability between \ac{SOME/IP} and \ac{TSN} by embedding \ac{QoS} signaling into existing \ac{SOME/IP} frameworks.
This integration simplifies the deployment of dynamic \ac{TSN} reservations within automotive \ac{SOA}.

A key limitation of current \ac{TSN} standards is the absence of standardized signaling procedures between applications, switches, and central controllers.
Our approach addresses this by embedding \ac{QoS} options in service negotiation, which existing \ac{SOME/IP} implementations ignore if unsupported, ensuring non-disruptive deployment.
The \ac{QoS} parameters transmitted during negotiation are tool-agnostic and can be adopted for other shapers/algorithms by \ac{TSN} central configuration tools (\eg for \ac{TAS} scheduling or \ac{ATS} parameter derivation) without modification.

All algorithms for determining idle slopes and worst-case latencies compared in this work are fully compatible with existing \ac{TSN} standards. 
The \codeword{TSN standard} inherently supports the centralized and decentralized configuration model but does not guarantee strict latency bounds, as shown in this paper.
The \codeword{delay budget} method is compatible with existing centralized \ac{TSN} configuration tools.
In~\cite{maile_decentral_2023}, we adapted the \codeword{delay budget} approach to decentralized stream reservation as well, \eg via \ac{RAP}, which requires changing the signaling to make every device aware of its own delay budgets. 

The complexity of our approach is favorable when compared to existing \ac{TSN} signaling methods.
\one~Signaling overhead is reduced by integrating \ac{QoS} negotiation directly into the \ac{SOME/IP} service discovery process,
eliminating the need for a two-stage procedure, \eg using \ac{SRP}~\cite{ieee8021q-22}.
This unification not only simplifies implementation but also minimizes the risk of inconsistent configurations that arise when multiple independent signaling mechanisms coexist.
\two~Operational complexity is lowered through centralization.
In general, centralized approaches are faster and less complex than distributed algorithms~\cite{ampbp-cacjr-20}, as no distributed agreement is required. 
\three~Algorithmic complexity is justified by performance: 
while the \ac{NC} implementation of the \codeword{delay budget} approach requires more sophisticated computation than basic \ac{TSN} formulas, it remains computationally efficient per reservation, and provides deterministic latency guarantees critical for \acp{IVN}.

\subsection{Does the Approach Converge in Dynamic Scenarios?}
\label{subsec:convergence}
Our approach ensures convergence in dynamic scenarios on three layers: 

\one Both \textbf{algorithms} for determining idle slopes and worst-case latencies are guaranteed to converge -- total flow analysis relying on delay budgets determines guarantees in fixed steps depending on the path length~\cite{maile_journal_2022}.

\two A \textbf{subscription} is established and bandwidth is reserved directly after acceptance -- utilizing the advantage of centralized configuration, for which no distributed algorithm needs to converge~\cite{ampbp-cacjr-20}.

\three Full \textbf{network} configuration converges -- the signaling scheme ensures that all established dynamic subscriptions are also reflected in the network configuration before acceptance is forwarded to respective subscribers.
However, no time guarantee can be provided for signaling reservations due to potential message loss. 
This behavior is consistent with other \ac{TSN} protocols, such as \ac{SRP}~\cite{ieee8021q-22}.
Retransmissions will ensure that the configuration is eventually applied.

Related work has shown in formal analysis that centralized \ac{TSSDN} can configure a typical automotive setup below \SI{50}{\milli\second}~\cite{te-asdts-16}.
As we do not propose new algorithms, detailed mathematical proof and convergence analysis are not within the scope of this paper, and we refer to related work, e.g., \cite{xu_convergence_PSO_2018} for the meta-heuristics, \cite{bujosa_srp_consistent_2021} for TSN signaling, and \cite{maile_decentral_2023} for the worst-case analysis. 

\subsection{What Are the Broader Implications for IVN Design?}
\label{subsec:ivn_implications}
While \ac{TSN} provides a unified backbone for \acp{IVN} with established static \ac{QoS} mechanisms, the traditional reservation model fails to accommodate dynamic services.
In modern vehicles, diverse and frequently updated software risks violations of safety-critical real-time requirements and increases configuration complexity. 

Our work addresses these limitations through a signaling scheme that integrates \ac{QoS} negotiation directly into automotive service discovery processes.
A centralized controller handles network reconfiguration transparently, which eliminates the need for involvement of end-device in reservation management -- a significant simplification compared to distributed approaches.

The design deliberately adopts only \ac{CBS} shaping in network bridges, explicitly excluding flow-based shaping, synchronous gating, or frame preemption mechanisms.
This strategic reduction in switch complexity aligns with other \ac{TSN}-based \ac{IVN} design proposals~\cite{wtm-avnjr-21}
The choice of CBS shifts complexity to idle slope calculation (a non-trivial problem, \cf Sec.~\ref{sec:problem}).

As confirmed by our simulations and prior work~\cite{b-adnjr-17,maile_decentral_2023}, worst-case latency analysis from \codeword{TSN standard} procedures cannot guarantee deadline-compliant reservations, particularly under topological variations. 
The \codeword{delay budget} approach overcomes this by leveraging \ac{NC} to rigorously derive idle slopes providing provable real-time guarantees~\cite{maile_decentral_2023} for dynamic configurations.

We adopt the \codeword{delay budget} method to dynamically configure a full \ac{IVN} for the first time.
Implementation reveals a fundamental trade-off: worst-case delay guarantees require substantial bandwidth over-reservation.
Empirical results show the \codeword{delay budget} method reserves approximately \SI{2.7}{\giga\bit\per\second}, 
over two time more than \codeword{CMI} (\SI{1.2}{\giga\bit\per\second}), and over 3 times more than naive traffic-load estimates (\codeword{flow interval} \SI{0.8}{\giga\bit\per\second}).
This disparity highlights the strong correlation between reserved bandwidth and worst-case latency,
particularly for slow flows with a tight deadline, that network designers must account for.

By combining signaling with rigorous mathematical modeling, we enable fully dynamic \ac{IVN} configuration that 
\one~maintains strict real-time compliance across diverse \ac{QoS} requirements, 
\two~scales to complex automotive topologies, and
\three~preserves compatibility with existing \ac{TSN} infrastructure.
This paradigm shift allows \acp{IVN} to support evolving software architectures while meeting strict timing requirements -- a capability previously limited to static configurations.

%% file: fig9_setuptime_cbsstudy.tex
\begin{tikzpicture}
    \begin{axis}[
        height=.33\linewidth,
        width=\linewidth,
        ylabel={Setup time},
        xlabel={Input links},
        x unit=\#,
        y unit=\milli\second,
        change y base=true,
        y SI prefix=milli,
        xmin = 2,
        xmax = 13,
        ymin = 0,
        ytick ={0,0.0012, 0.001, 0.0008, 0.0006, 0.0004, 0.0002},
        xtick ={2, 3, 4, 5, 6, 7, 8, 9, 10, 11, 12, 13},
        xtick pos=bottom,
        legend style={
            rounded corners,
            at={(0.825,0.1)}, 
            legend cell align={left},
            anchor=south, 
            legend columns=1, 
        },
    ]
    \addplot [very thick, no markers, CoreGreen] table [x=numInputLinks, y=S1, col sep=comma] {setupTimeCMI_PCT.csv};
    \addplot [very thick, no markers, CoreBlue] table [x=numInputLinks, y=S7, col sep=comma] {setupTimeCMI_PCT.csv};
    \addplot [very thick, no markers, CoreRed] table [x=numInputLinks, y=S15, col sep=comma] {setupTimeCMI_PCT.csv};
    \legend{
        {1 Stage},
        {7 Stages},
        {15 Stages},
        }
    \end{axis}
\end{tikzpicture}

%% file: fig10_idleslopeconfig.tex
\begin{tikzpicture}
    \begin{axis}[
        height=.35\linewidth,
        width=\linewidth,
        ylabel={Idle slope},
        xlabel={Input links},
        x unit=\#,
        y unit=Mbit/s,
        change y base=true,
        y SI prefix=mega,
        xmin = 2,
        xmax = 13,
        ymin = 0,
        ymax = 105000000,
        ytick ={0, 25000000, 50000000, 75000000},
        xtick ={2, 3, 4, 5, 6, 7, 8, 9, 10, 11, 12, 13},
        xtick pos=bottom,
        legend style={
            rounded corners,
            at={(0.015,0.97)}, 
            legend cell align={left},
            anchor=north west,
            legend columns=5,
        },
    ]
    \addplot [very thick, no markers, CoreBlue] table [x=numInputLinks, y=CMI, col sep=comma] {idleSlopeConfig5Stages.csv};%
    \addplot [very thick, no markers, CoreRed] table [x=numInputLinks, y=SI, col sep=comma] {idleSlopeConfig5Stages.csv};%
    \addplot [very thick, no markers, CoreGreen] table [x=numInputLinks, y=NC_Avg, col sep=comma] {idleSlopeConfig5Stages.csv};%
    \addplot [very thick, no markers, CoreGreen, dashed] table [x=numInputLinks, y=NC_Min, col sep=comma] {idleSlopeConfig5Stages.csv};%
    \addplot [very thick, no markers, CoreGreen, dotted] table [x=numInputLinks, y=NC_Max, col sep=comma] {idleSlopeConfig5Stages.csv};%
    \legend{
            CMI,
            FI,
            DB Avg,
            DB Min,
            DB Max
        }%
    \end{axis}%
\end{tikzpicture}%

%% file: fig11_aggdelayq.tex
\begin{tikzpicture}
    \begin{axis}[
        height=.33\linewidth,
        width=\linewidth,
        ylabel={Queue delay},
        xlabel={Stages},
        x unit=\#,
        y unit=\milli\second,
        change y base=true,
        y SI prefix=milli,
        xmin = 1,
        xmax = 15,
        ymin = 0,
        ymax = 0.0035,
        ytick ={0, 0.001, 0.002, 0.003},
        xtick ={1, 2, 3, 4, 5, 6, 7, 8, 9, 10, 11, 12, 13, 14, 15},
        xtick pos=bottom,
        legend style={
            rounded corners,
            at={(0.015,0.97)}, 
            legend cell align={left},
            anchor=north west,
            legend columns=3,
        },
    ]
    \addplot [very thick, no markers, CoreBlue] table [x=numStages, y=Sim_CMI_BECT, col sep=comma] {aggregateDelayWithQ4IL.csv};
    \addplot [very thick, no markers, CoreBlue, dashed] table [x=numStages, y=Sim_CMI_PCT, col sep=comma] {aggregateDelayWithQ4IL.csv};
    \addplot [very thick, no markers, CoreBlue, dotted] table [x=numStages, y=Q_CMI, col sep=comma] {aggregateDelayWithQ4IL.csv};
    \addplot [very thick, no markers, CoreRed] table [x=numStages, y=Sim_SI_BECT, col sep=comma] {aggregateDelayWithQ4IL.csv};
    \addplot [very thick, no markers, CoreRed, dashed] table [x=numStages, y=Sim_SI_PCT, col sep=comma] {aggregateDelayWithQ4IL.csv};
    \addplot [very thick, no markers, CoreRed, dotted] table [x=numStages, y=Q_SI, col sep=comma] {aggregateDelayWithQ4IL.csv};
    \legend{
            Sim CMI BECT,
            Sim CMI PCT,
            Q-WC CMI,
            Sim FI BECT,
            Sim FI PCT,
            Q-WC FI,
        }
    \end{axis}
\end{tikzpicture}%

%% file: fig12_study_e2e_combined.tex
\begin{tikzpicture}
    \begin{groupplot}[
        group style={
            group size=1 by 2,
            horizontal sep = 0pt, 
            vertical sep = 0pt,
            xlabels at = edge bottom,
            xticklabels at = edge bottom, 
            ylabels at = edge left,
            yticklabels at = edge left, 
        },
        height=.3\linewidth,
        width=\linewidth,
        xlabel={Stages},
        x unit=\#,
        change y base=true,
        y SI prefix=milli,
        xmin = 1,
        xmax = 15,
        xtick ={1, 2, 3, 4, 5, 6, 7, 8, 9, 10, 11, 12, 13, 14, 15},
        ytick ={0, 0.005, 0.01, 0.015},
        ytick pos=left,
        xtick pos=bottom,
        ymin = 0,
        ymax = 0.017,
    ]
    \nextgroupplot[
        ylabel={E2E delay},     
        y unit=\milli\second,   
        y label style={at={(axis description cs:-0.03,0)},anchor=south},
        legend style={
        rounded corners,
        at={(0.015,0.97)}, 
        legend cell align={left},
        anchor=north west,
        legend columns=3,
    }]
    \addplot [very thick, no markers, CoreBlue] table [x=numStages, y=CMI_BE, col sep=comma] {e2eDelaysIL13.csv};
    \addplot [very thick, no markers, CoreBlue, dashed] table [x=numStages, y=CMI_PCT, col sep=comma] {e2eDelaysIL13.csv};
    \addplot [very thick, no markers, CoreBlue, dotted] table [x=numStages, y=QCMI, col sep=comma] {e2eDelaysIL13.csv};
    \addplot [very thick, no markers, CoreRed] table [x=numStages, y=SI_BE, col sep=comma] {e2eDelaysIL13.csv};
    \addplot [very thick, no markers, CoreRed, dashed] table [x=numStages, y=SI_PCT, col sep=comma] {e2eDelaysIL13.csv};
    \addplot [very thick, no markers, CoreRed, dotted] table [x=numStages, y=QSI, col sep=comma] {e2eDelaysIL13.csv};
    \legend{
        Sim CMI BECT,
        Sim CMI PCT,
        Q-WC CMI,
        Sim FI BECT,
        Sim FI PCT,
        Q-WC FI,
    }

    \nextgroupplot[
        legend style={
            rounded corners,
            at={(0.015,0.97)}, 
            legend cell align={left},
            anchor=north west,
            legend columns=2,
        },
    ]
        \addplot [very thick, no markers, CoreGreen] table [x=numStages, y=NC_BE, col sep=comma] {e2eDelaysIL13.csv};
        \addplot [very thick, no markers, CoreGreen, dashed] table [x=numStages, y=NC_PCT, col sep=comma] {e2eDelaysIL13.csv};
        \addplot [very thick, no markers, CoreMagenta, dotted] table [x=numStages, y=DB, col sep=comma] {e2eDelaysIL13.csv};
        \addplot [very thick, no markers, CoreGreen, dotted] table [x=numStages, y=DB_INDEPENDENT, col sep=comma] {e2eDelaysIL13.csv};
        \legend{
            Sim DB BECT,
            Sim DB PCT,
            DB-WC current,
            DB-WC independent,
        }
    \end{groupplot}
    \fill[very thick, pattern=north west lines, pattern color=LightCoreRed] (12.75,-0.01) rectangle (14.853,-3.345);
    \node[anchor=north west, align=center, rotate=90] at (13.28,-3.0) {no feasible \\ schedule found};
\end{tikzpicture}

%% file: fig12_car_idleslopeconfig.tex
\begin{tikzpicture}
    \begin{axis}[
        height=.4\linewidth,
        width=\linewidth,
        ylabel={Idle slope},
        xlabel={Simulation time},
        x unit=\milli\second,
        change x base=true,
        x SI prefix=milli,
        y unit=Gbit/s,
        change y base=true,
        y SI prefix=giga,
        xmin = 0.0907200085,
        xmax = 0.101053191806,
        ymin = 0,
        ymax = 2500000000,
        ytick ={0,500000000, 1000000000, 1500000000, 2000000000, 2500000000},
        xtick ={0.091, 0.092, 0.093, 0.094, 0.095, 0.096, 0.097, 0.098, 0.099, 0.1, 0.101},
        xtick pos=bottom,
        legend style={
            rounded corners,
            at={(0.015,0.97)}, 
            legend cell align={left},
            anchor=north west, 
            legend columns=3,
        },
        const plot,
        legend image post style={sharp plot},
    ]
    \addplot [very thick, no markers, const plot, CoreBlue, dotted] table [x=CMI_B250_4_time, y=CMI_B250_4_value, col sep=comma] {switchCenter-2_idleslopes.csv};
    \addplot [very thick, no markers, const plot, CoreBlue, dashed] table [x=CMI_B250_5_time, y=CMI_B250_5_value, col sep=comma] {switchCenter-2_idleslopes.csv};
    \addplot [very thick, no markers, const plot, CoreBlue] table [x=CMI_B250_Total_time, y=CMI_B250_Total_value, col sep=comma] {switchCenter-2_idleslopes.csv};
    \addplot [very thick, no markers, const plot, CoreRed, dotted] table [x=SI_4_time, y=SI_4_value, col sep=comma] {switchCenter-2_idleslopes.csv};
    \addplot [very thick, no markers, const plot, CoreRed, dashed] table [x=SI_5_time, y=SI_5_value, col sep=comma] {switchCenter-2_idleslopes.csv};
    \addplot [very thick, no markers, const plot, CoreRed] table [x=SI_Total_time, y=SI_Total_value, col sep=comma] {switchCenter-2_idleslopes.csv};
    \addplot [very thick, no markers, const plot, CoreGreen, dotted] table [x=Dyrectsn_4_time, y=Dyrectsn_4_value, col sep=comma] {switchCenter-2_idleslopes.csv};
    \addplot [very thick, no markers, const plot, CoreGreen, dashed] table [x=Dyrectsn_5_time, y=Dyrectsn_5_value, col sep=comma] {switchCenter-2_idleslopes.csv};
    \addplot [very thick, no markers, const plot, CoreGreen] table [x=Dyrectsn_Total_time, y=Dyrectsn_Total_value, col sep=comma] {switchCenter-2_idleslopes.csv};
    \legend{
            CMI PCP 5,
            CMI PCP 6,
            CMI PCP 5 $+$ 6,
            FI PCP 5,
            FI PCP 6,
            FI PCP 5 $+$ 6,
            DB PCP 5,
            DB PCP 6,
            DB PCP 5 $+$ 6
        }
    \end{axis}
\end{tikzpicture}%

%% file: tab3_car_combined_is_e2e.tex
\begin{tabularx}{\linewidth}{l l Y Y Y Y Y Y}
    \toprule
     &  & \multicolumn{3}{c}{\textbf{Idle slope} [\si{\mega\bit\per\second}]} & \multicolumn{3}{c}{\textbf{E2E delay} [\si{\milli\second}]} \\
    \textbf{PCP} & \textbf{Reservation scheme} & \textbf{Min} & \textbf{Avg} & \textbf{Max} & \textbf{Min} & \textbf{Avg}  & \textbf{Max} \\
    \midrule
    5  & class measurement interval & $21$ & $244$ & $398$ & $0.01$ & $0.08$ & $0.30$ \\
    (CAN) & flow interval & $<1$ & $<1$ & $1$  & $0.01$ & \textcolor{red}{$23.46$} & \textcolor{red}{$148$} \\
    & delay budget & $24$ & $299$ & $512$ & $0.01$ & $0.08$ & $0.28$ \\
    \midrule
    6 & class measurement interval & $129$ & $480$ & $886$ & $0.04$ & $0.09$ & $0.17$ \\
    (Camera, & flow interval & $107$ & $428$ & $785$ & $0.04$ & $0.11$ & $0.21$ \\
    Lidar) & delay budget & $194$ & $1004$ & $2212$ & $0.03$ & $0.06$ & $0.12$ \\
    \bottomrule
\end{tabularx}%

%% file: 7_conclusion.tex
\section{Conclusion and Outlook} 
\label{sec:conclusion_and_outlook}
Critical dynamic in-vehicle services require hard real-time guarantees, which must be configured within the automotive SOA. 
We identified three key challenges that hinder deadline-compliant operation of dynamic services, in particular the lack of a standardized mechanism to communicate \ac{QoS} requirements of automotive services in a TSN.

To address these challenges, we proposed an integrated \ac{QoS} signaling scheme within the automotive SOA, enabling service providers, consumers, and the network to negotiate \ac{QoS} requirements. 
We use the OpenFlow protocol to intercept the automotive SOA protocol and collect the \ac{QoS} demands. 
With this information, a central \ac{TSSDN} controller can reserve resources and guarantee latency bounds for critical services, for which we compare the TSN standard approach with a fixed CMI and based on flow intervals against a delay budget approach.

This work closes the research gap  of configuring CBS idle slopes for guaranteed latency bounds in  
 dynamic service environments at the full complexity of an IVN, which has not been possible so far.
For this we compared bandwidth reservation schemes and their impact on resource allocation and network delay, revealing significant differences for realistic use cases. 
In particular, we identified counterexamples to delay bounds derived from TSN standard worst-case analysis, reinforcing previous findings~\cite{b-adnjr-17,maile_decentral_2023}.

Our analysis demonstrated that the delay budget approach is the only method for correctly implementing hard latency bounds. 
In a full \ac{IVN} case study, our service negotiation framework successfully configured 450 subscriptions within just \SI{11.1}{\milli\second}, showcasing the feasibility of our integrated solution. 
The combination of our \ac{TSSDN} signaling scheme with the delay budget approach effectively supports dynamic real-time communication in automotive networks.

Future signaling extensions should adapt the signaling approach to additional SOA protocols, such as DDS. 
We plan to incorporate additional parameters, \eg to signal security requirements for authentication and authorization~\cite{sms-pdcta-25}.
Further optimizing delay budget selection shall enhance service scheduling and latency guarantees. 
Here, \ac{DRL} and \ac{ML} methods could be explored to further improve scheduling and resource allocation, potentially leading to more efficient and adaptive network management.
Finally, our solution could be extended to support other TSN mechanisms, including \ac{ATS} (IEEE 802.1Qcr), path redundancy (IEEE 802.1CB), and ingress control (IEEE 802.1Qci). 
We further propose to investigate the dynamic adaptation of delay budgets after their initial assignment, as well as to develop more abstract mathematical proofs of the total system performance, \eg using formal methods such as contract theory~\cite{graf_contract_theory_2018}.